\documentclass[manuscript]{elsarticle}
\usepackage[gen]{eurosym}
\usepackage{graphicx}
\graphicspath{ {images/} }
\usepackage{amsmath}
\usepackage[margin=1in,marginpar=1in]{geometry}
\usepackage{tabularx}
	\newcolumntype{Z}{>{\centering\arraybackslash}p}
\usepackage{setspace}
\usepackage[usenames, dvipsnames]{color}
\usepackage[normalem]{ulem}
\usepackage{ todonotes}
\usepackage{siunitx}
\usepackage{url}

\journal{arXiv}

\begin{document}
	
\begin{frontmatter}
		
\title{Digital twin of a MWh-scale grid battery system for efficiency and degradation analysis}
\author[OU]{Jorn M. Reniers} 
	\ead{jorn.reniers@eng.ox.ac.uk}
\author[OU]{David A. Howey \corref{cor1}} 
	\ead{david.howey@eng.ox.ac.uk}
	\ead[url]{http://howey.eng.ox.ac.uk/}

\address[OU] {Department of Engineering Science, University of Oxford, OX1 3PJ, Oxford, UK}
\cortext[cor1]{Corresponding author}

\begin{abstract}
    Large-scale grid-connected lithium-ion batteries are increasingly being deployed to support renewable energy roll-out on the power grid. These battery systems consist of thousands of individual cells and various ancillary systems for monitoring and control. Although many studies have focused on the behaviour of single lithium-ion cells, the impact of system design choices and ancillary system controls on long-term degradation and efficiency of system, containing thousands of cells, has rarely been considered in detail. Here, we simulate a \SI{1}{MWh} grid battery system consisting of 18900 individual cells, each represented by a separate electrochemical model, as well as the thermal management system and power electronic converters. Simulations of the impact of cell-to-cell variability, thermal effects, and degradation effects were run for up to 10000 cycles and 10 years. It is shown that electrical contact resistances and cell-to-cell variations in initial capacity and resistance have a smaller effect on performance than previously thought. Instead, the variation in degradation rate of individual cells dominates the system behaviour over the lifetime. The importance of careful thermal management system control is demonstrated, with proportional control improving overall efficiency by 5 \%-pts over on-off methods, also increasing the total usable energy of the battery by 5 \%-pts after 10 years.
	\textit{keywords}: lithium-ion battery, degradation, efficiency, digital twin
\end{abstract}

	% Results from 10-year simulation, see near the end of the manuscript
	% mean capacities at end of 10 years:
	%   1   81.1897174933862
	%   2   73.9822480820106
	%   3   75.6710346395503
	%   4   76.5815488988095
	%   5   77.9584026851852
	
    % Overall efficiency at start, 10 years, difference
    % 67.8154624215420	60.6725850372817	-7.14287738426030   note initial steep drop
    % 80.2319549340633	66.2573220880796	-13.9746328459837   note alternating
    % 77.8406472706355	69.1010005196294	-8.73964675100619   note alternating
    % 81.9471141767527	77.0867772983265	-4.86033687842624
    % 81.4637760702745	75.4151776513717	-6.04859841890286
	
	% Total usable energy
	% start             end                 difference
    % 82.3488845988846	57.8398398398398	-24.5090447590448
    % 86.2449860574861	49.1260099385099	-37.1189761189761
    % 84.9305734305734	52.1503181753182	-32.7802552552553
    % 86.1036299692550	54.4675809738310	-31.6360489954240
    % 85.8544017231517	54.1397647647648	-31.7146369583870

\end{frontmatter}

\newpage 

\section{Introduction}

    Traditionally, fluctuations in electricity generation and demand were met by flexible generation units and hydro storage. The rapid deployment of renewable energy generators increases the need for grid balancing, voltage support, and other services \cite{Solomon2014}. Due to rapid cost declines of lithium-ion batteries \cite{Nykvist2019, ziegler2021re}, they are increasingly becoming an important part of grid infrastructure, participating in the various markets for frequency control, system reserves, and wholesale energy trading \cite{Gunter2016}.

    Grid-connected lithium-ion batteries are large, complex systems consisting of thousands of cells and various ancillary systems. Power electronic converters create an AC voltage and current from the variable DC battery pack voltage, a thermal management system ensures stable temperatures, an energy management system handles the high-level system control, and lower-level battery management systems monitor individual cells to ensure safety \cite{Hesse2017}. 

    Recently, models have been designed to simulate the various grid battery components and their interactions. Patsios et al.\ \cite{Patsios2016} used detailed models for the transformer, power converter, and a battery cell. They found that losses in the power electronic converter outweigh losses in the cells, and that the control system needs to trade-off efficiency and degradation since operating the battery at low state-of-charge (SoC) typically reduces degradation but increases energy losses. Schimpe et al.\ \cite{Schimpe2018c} used simpler models, but accounted for many different components and the thermal interactions within the battery. They investigated the different sources of losses in a \SI{192}{kWh} system and found that the converters and ancillary systems dominated the losses, especially at low utilisation. They did not include battery degradation in their simulation.

    Both Patsios et al.\ and Schimpe et al.\ simulated a single battery cell and multiplied its current and voltage with the number of parallel- and series-connected cells in the system to obtain an estimate of the behaviour of the entire battery pack. This assumes that all cells behave identically, while in reality it has been shown that even cells from the same production batch are slightly different to one another. Barreras et al.\ \cite{Barreras2017a} screened over 200 cells and found that the standard deviations for the initial capacity and resistance were 0.4\% and 2.5\% respectively. As cells degrade, these differences increase---when Harris et al.\ \cite{Harris2017} aggressively cycled 24 cells, the `worst' cell only had 45\% remaining capacity while the `best' cell still had 85\% at the end of the experiment. Similarly, Baumh$\ddot{o}$fer et al.\ \cite{Baumhofer2014} cycled 48 cells and found that when the mean cell capacity had decreased to about 70\% of nominal, the `worst' and `best' cells had a remaining capacity of about 60\% and 80\% respectively.

    Very few pack simulation studies include cell-to-cell variations due to the associated computational challenges of simulating every single cell. Rumpf et al.\ \cite{Rumpf2018} used a detailed multiphysics model to explore how the current in a small parallel module is divided inhomogeneously between the cells due to interconnection and contact resistances and differences between the cells. Liu et al.\ \cite{Liu2019} predicted that for a medium-sized module, these effects may reduce the accessible energy by 6\%, and that additionally the degradation rate may increase by about 5\%. Dubarry et al.\ \cite{Dubarry2019a} took a different approach and combined a number of empirical models, each simulating a different aspect of the behaviour of the system, and were able to simulate a small battery consisting of a few hundred cells. Finally, Rogers et al.\ \cite{Rogers2021} did not explicitly simulate the  cells, but instead considered the statistical distribution of cell capacities in large-scale batteries and the impact of battery configuration on usable capacity. Notably, none of these papers account for the ancillary components of a grid-scale battery. In addition, none of the physics-based or empirical approaches were scalable up to simulations of thousands of cells, although one solution to this is the statistical approach of Rogers et al.\

    Here, we use a multi-physics approach to simulate a MWh-size battery, consisting of almost twenty thousand cells, over its entire plausible lifetime. Every battery cell was simulated individually, with its own physics-based model including degradation and thermal effects, allowing direct consideration of the impacts of cell-to-cell variations in parameters. Various ancillary systems and their interactions with the cells were included in the overall system simulation.

    Regarding nomenclature, different levels within the battery system are named here according to the convention of Schimpe et al.\ \cite{Schimpe2018c}, i.e.\ the lowest level are called `cells', a group of parallel-connected cells (e.g.\ 7 parallel cells) is called a `block', a group of blocks in series is a `module' (e.g.\ 20 of the aforementioned blocks connected in series would be a 20s7p module), a group of modules is a `rack' (e.g.\ 15 series-connected modules would be a rack of 15s20s7p cells), a group of racks is a `battery compartment' (e.g.\ 9 parallel racks would be described as 9p15s20s7p cells), and finally a `container' consists of the battery compartment and the ancillary systems. A `unit' can be any of the levels.

    In the subsequent sections we introduce the hierarchical modelling approach used in this paper, and then give results and discussion of various system simulations exploring the impact of cell-to-cell variations, electrical contact resistances, and thermal management system design.

\section{Methods}

	In this section we explain the various models used to simulate the full battery system. The equations for the battery cell models (including degradation), plus thermal, electrical and ancillary systems are given, and relevant implementation issues are discussed in each sub-section.
	
	The overall battery system model was implemented in C++. Object-oriented programming was used to ensure modularity and impose the correct hierarchy by having each class be responsible for its own sub-models. The \emph{Cell} class implements all equations for the battery cell model, including computing the cell's voltage, temperature and degradation state based on the applied current and thermal conditions. The \emph{Group} class implements all the functionality needed to join cells, blocks, modules etc.\ electrically in series or parallel. It takes advantage of the polymorphism of object-oriented programming to work with any units, such that it may be used to simulate blocks, modules, racks and battery compartments. The Group class is responsible for the electrical and thermal interactions between the units connected to it.

	\subsection{Cell model including degradation} \label{SPM}
		The single particle model (SPM) was used to simulate individual Li-ion cells \cite{Ning2004}. It is a basic electrochemical battery model where electrolyte ionic transport is assumed to be much faster than transport within the solid electrodes, such that it can be neglected. Each electrode is be simulated by a single `average' spherical domain of active material within which lithium ions diffuse according to Fick's law, with temperature-dependent diffusion constant $D_i(T)$ (Equation \ref{eqn:SPM_diffusion}) \cite{Marquis2019}. For boundary conditions, in the respective electrode $i$, the diffusion at the centre must be zero due to symmetry (Equation \ref{eqn:SPM_BCcentre}), while at the particle surface, the concentration gradient must be compatible with the intercalating lithium flux $j_i$, which is proportional to the current density $i_i$, and therefore to the overall applied cell current $I$ (Equation \ref{eqn:SPM_BCsurface}). Here, the product of the electrode thickness $\tau_i$, the electrode geometric surface area $A_i$, and the effective surface area (which is a function of the volume fraction of active material $\epsilon_i$ and the particle radius $R_i$), gives the scaling factor representing the `amount of active material' in an electrode. Initial conditions were assumed to be temperature uniform and ambient, and uniform concentration giving an overall initial state of charge of 50\%.
		\begin{equation} % diffusion PDE
		\label{eqn:SPM_diffusion}  
		\frac{\partial c_i(r)}{\partial t} = \frac{D_i(T)}{r^2}\frac{\partial}{\partial r} \left( r^2\frac{\partial c_i(r)}{\partial r} \right) 
		\end{equation}
		\begin{equation} % diffusion PDE boundary condition at the centre
		\label{eqn:SPM_BCcentre}
		D_i(T) \frac{\partial c_i(r)}{\partial r} \bigg\rvert _{r = 0} = 0 
		\end{equation}
		\begin{equation} % diffusion PDE boundary condition at the surface
		\label{eqn:SPM_BCsurface}
		D_i(T) \frac{\partial c_i(r)}{\partial r} \bigg\rvert _{r = R_i} = \pm j_i = \pm \frac{i_i}{nF}  = \pm \frac{I}{nF 3 \frac{\epsilon_i}{R_i} A_i \tau_i} 
		\end{equation}
		
		The lithium intercalation reaction at the surface of an electrode follows Bulter-Volmer kinetics (Equation \ref{eqn:SPM_BV}) with overpotential $\eta_i$, transfer coefficient $\alpha$, gas constant $R$, and temperature $T$. The exchange current density $i_{0,i}$ is a function of the temperature-dependent rate constant $k_i(T)$, maximum lithium concentration $c_i^{ \text{max}}$ and lithium concentration in the electrolyte $c_{ \text{el}}$, Equation \ref{eqn:SPM_i0}:
		\begin{equation}   % bulter-volmer equation for li insertion
		\label{eqn:SPM_BV}
		i_i = i_{0,i} \left( \text{exp} \left( -\frac{\alpha nF}{R T} \eta_i \right) -  \text{exp} \left( \frac{(1-\alpha) nF}{R T} \eta_i \right) \right)  
		\end{equation}
		\begin{equation} % exchange current density of the li intercalation
		\label{eqn:SPM_i0}
		i_{0,i} = nFk_i\left(T\right)c_i\left(R_i\right)^\alpha c_{ \text{el}}^{1-\alpha} \left( c_i^{ \text{max}}-c_i\left(R_i\right) \right)^{1-\alpha} 
		\end{equation} 
		
		A bulk energy balance (Equation \ref{eqn:SPM_temp}) governs the lumped temperature changes of the cell with density $\rho_{\text{cell}}$, surface area $A_{\text{cell}}$, thickness $\tau_{\text{cell}}$ and heat capacity $C_{p,\text{cell}}$ \cite{Guo2011}. Internal heat generation originates from ohmic heating with internal resistance $R_{\text{dc,cell}}$, reaction heating from the intercalation reactions with overpotentials $\eta_i$, and entropic heating, represented by the entropic coefficient $\frac{\partial U_{\text{cell}}}{\partial T}$. The heat exchange of a cell with all neighbouring elements is calculated using the cell temperature and heat transfer coefficient between the cell and each neighbouring element, as explained in section \ref{thermalCoupling}. Arrhenius relations with activation energy $E$ were used to simulate temperature-dependent parameters, where $X$ is either the diffusion constant or rate constant (Equation \ref{eqn:SPM_arrhenius}):
		\begin{equation} % bulk thermal model ODE
		\label{eqn:SPM_temp}
		\rho_{\text{cell}} A_{\text{cell}} \tau_{\text{cell}} C_{p,\text{cell}} \frac{\partial T}{\partial t} = \: I^2 R_{\text{dc,cell}} + I \left( \eta_{n} - \eta_{p} \right) + IT\frac{\partial U_{\text{cell}}}{\partial T} - \sum_l \left( h_l A_{\text{cell}} \left( T - T_l \right) \right) 
		\end{equation}
		\begin{equation} % Arrhenius relation for diffusion constant
		\label{eqn:SPM_arrhenius}
		X(T)=X^{\text{ref}} \text{exp}\left[- \frac{E_{X}}{R}  \left( \frac{1}{T} - \frac{1}{T^{\text{ref}}} \right) \right] 
		\end{equation} 
		
		Two degradation models were used to simulate the decrease in charge capacity and increase in resistance of cells as they age. Firstly, the growth of a passivation layer on the graphite, known as the solid electrolyte interphase (SEI) layer, consumes lithium. The Christensen and Newman model \cite{Christensen2005} of solvent reduction was used to simulate this process (Equation \ref{eqn:SEI}). This assumes temperature-dependent diffusion with a linear concentration gradient through the existing SEI layer of thickness $\tau_{\text{sei}}$ and diffusion constant $D_{\text{sei}}(T)$, and Tafel reaction kinetics with temperature-dependent rate constant $k_{\text{sei}}(T)$, anode potential $U_n$, anode overpotential $\eta_n$ and potential of the SEI reaction $U_{\text{sei}}$. The SEI layer thickness growth rate increases in proportion to the SEI current density, increasing the cell resistance, and the side reaction current density is added to the surface boundary condition on the anode (Equation \ref{eqn:SEI_LLI}), removing lithium from the cell. The growing SEI layer also clogs pores, reducing the amount of active material that can be accessed by the intercalating lithium. To capture this effect, the model of Ashwin et al.\ \cite{Ashwin2016} was used, equation \ref{eqn:SEI_pore}---this assumes a linear correlation between the loss of active material and the weighted sum of SEI current density and main intercalation current density, with partial molar volumes denoted $v$, and a fitting constant $\beta_1$.
		\begin{equation}
		\label{eqn:SEI}
		i_{\text{sei}} = \frac{\text{exp} \left( -\frac{\alpha_{\text{sei}} nF}{R T} \eta_{\text{n}}  \right)   } { \frac{1}{nFk_{\text{sei}}(T) \text{exp} \left(  -\frac{\alpha_{\text{sei}} nF}{R T} \left( U_{\text{n}} - U_{\text{sei}} \right)  \right) } + \frac{\tau_{\text{sei}}}{nF D_{\text{sei}}(T)} } 
		\end{equation}
		\begin{equation} % SEI diffusion PDE boundary conditon for LLI
		\label{eqn:SEI_LLI}
		D_{\text{n}}(T) \frac{\partial c_{\text{n}}(r)}{\partial r} \bigg\rvert _{r = R_n} = - \frac{i_{\text{n}}}{nF} - \frac{i_{\text{sei}}}{nF}
		\end{equation}
		\begin{equation} % SEI pore clogging
		\label{eqn:SEI_pore}
		\frac{\partial \epsilon_{\text{n}}}{\partial t} = - \beta_1 \left( v_{\text{sei}} i_{\text{sei}} + v_{\text{li}} i_n \right)
		\end{equation}
		
		Although SEI growth is often considered the main degradation mechanism in Li-ion cells, models to simulate it cannot predict an intrinsic `knee point' or `roll-over point', i.e.\ when degradation rate suddenly increases later in life \cite{Reniers2019a}. Because cell-to-cell variations have been shown to rapidly increase beyond this point \cite{Baumhofer2014}, a second degradation model was also included, simulating loss of active material (LAM) in the electrodes. In this model, alternating stresses due to successive (de)intercalation cycles cause crack growth, which can electrically isolate parts of the active material. To simulate this, a crack growth model driven by stresses in the materials was used. Dai et al.\ \cite{Dai2014} derived equations for the radial (equation \ref{eqn:stress_r}), tangential (equation \ref{eqn:stress_t}) and hydrostatic (equation \ref{eqn:stress_h}) stresses in spherical particles with lithium concentration gradients, where $\Omega$ is the partial molar volume, $Y$ is the Young's modulus in, $\nu$ is the Poisson's ratio and $\zeta$ is a dummy integration variable, as follows:
		\begin{equation} % Radial stress according to Dai et al
		\label{eqn:stress_r}
		\sigma_{r,i}(r) = \frac{2 \Omega_i Y_i}{3 \left( 1 - \nu_i \right)} \left( \frac{1}{R_i^3} \int_{0}^{R_i} c_i(r) r^2 \text{d}r - \frac{1}{r^3} \int_{0}^{r} c_i(\zeta) \zeta^2 \text{d}\zeta    \right)
		\end{equation}
		\begin{equation} % Tangential stress according to Dai et al
		\label{eqn:stress_t}
		\sigma_{t,i}(r) = \frac{ \Omega_i Y_i}{3 \left( 1 - \nu_i \right)} \left( \frac{2}{R_i^3} \int_{0}^{R_i} c_i(r) r^2 \text{d}r + \frac{1}{r^3} \int_{0}^{r} c_i(\zeta) \zeta^2 \text{d}\zeta   - c_i(r,t)  \right)
		\end{equation}
		\begin{equation} % Hydrostatic stress
		\label{eqn:stress_h}
		\sigma_{h,i}(r) = \frac{\sigma_{r,i}(r) + 2 \sigma_{t,i}(r) }{3}
		\end{equation}
		
		Very few models are available to link stress to loss of active material. Most models are designed for crack growth within the SEI layer. However, because the underlying mechanisms are the same, such a model was assumed here to simulate a reduction in the volume fraction of active material. The crack growth model from Laresgoiti et al.\ \cite{Laresgoiti2015} is based on crack growth in metals, and assumes that cracks grow in proportion to the difference between the maximum and minimum stresses per cycle, respectively $\sigma_{h,i}^{\text{max}}$ and  $\sigma_{h,i}^{\text{min}}$, normalised by the yield strength $\sigma_{\text{yield},i}$, and raised to the power $1/m$, Equation \ref{eqn:stress_LAM}. A constant $\beta_2$ relates the crack growth rate to LAM. 
		\begin{equation} % Dai + Laresgoiti's LAM model
		\label{eqn:stress_LAM}
		\frac{\partial \epsilon_i}{\partial t} = \beta_2 \left( \frac{\sigma_{h,i}^{\text{max}} - \sigma_{h,i}^{\text{min}}}{\sigma_{\text{yield},i}} \right) ^{\frac{1}{m}}
		\end{equation}
		
		In summary, cyclable lithium is removed irreversibly due to SEI growth, and available active material is reduced due to the crack growth. Both mechanisms also increase the total resistance of the cell $R_{\text{dc,cell}}$ (Equation \ref{eqn:SPM_Rdc}), where $r_{\text{dc},i}$ is respectively the specific resistance of the anode, cathode and SEI layer. The specific resistances are divided by the total surface area, such that a reduction in active material will increase the overall resistance. The resistance of the SEI layer is relative to its volume, where the total surface area is the same as the anodic surface area and the required thickness is the thickness of the SEI layer.		
		\begin{equation} % Cell DC resitance
		\label{eqn:SPM_Rdc}
		R_{\text{dc,cell}} =\frac{r_{\text{dc},n}}{3 \frac{\epsilon_n}{R_n} A_n \tau_n} + \frac{r_{\text{dc},p}}{3 \frac{\epsilon_p}{R_p} A_p \tau_p} + \frac{r_{\text{dc,sei}}}{3 \frac{\epsilon_n}{R_n} A_n \tau_n}\tau_{\text{sei}}
		\end{equation}
		
		The measurable terminal voltage of the cell is given by Equation \ref{eqn:SPM_V}, where $U_i^{\text{ref}}$ is the open circuit potential of electrode $i$ (positive or negative) at a standard reference temperature and at respective surface concentration $c_i$; the entropic contribution to potential is given by $\frac{\partial U_{\text{cell}}}{\partial T}$, which is a function of SOC, the kinetic overpotentials are $\eta_i$ respectively, and the ohmic resistance voltage drop is $R_{\text{dc},\text{cell}} I_{\text{cell}}$: 
		\begin{equation}
		\label{eqn:SPM_V}
		V_{\text{cell}} =  U_p^{\text{ref}}\left( c_p\left( R_p\right)  \right) - U_n^{\text{ref}}\left( c_n\left( R_n\right)  \right) + \left( T-T^{\text{ref}} \right) \frac{\partial U_{\text{cell}}}{\partial T}  - \left( \eta_{n}-\eta_p \right) - R_{\text{dc},\text{cell}} I_{\text{cell}} 
		\end{equation}
		
		\subsection{Model parameterisation} \label{sec:parameterisation}
\begin{figure}
			\centering
			\includegraphics[width=8cm]{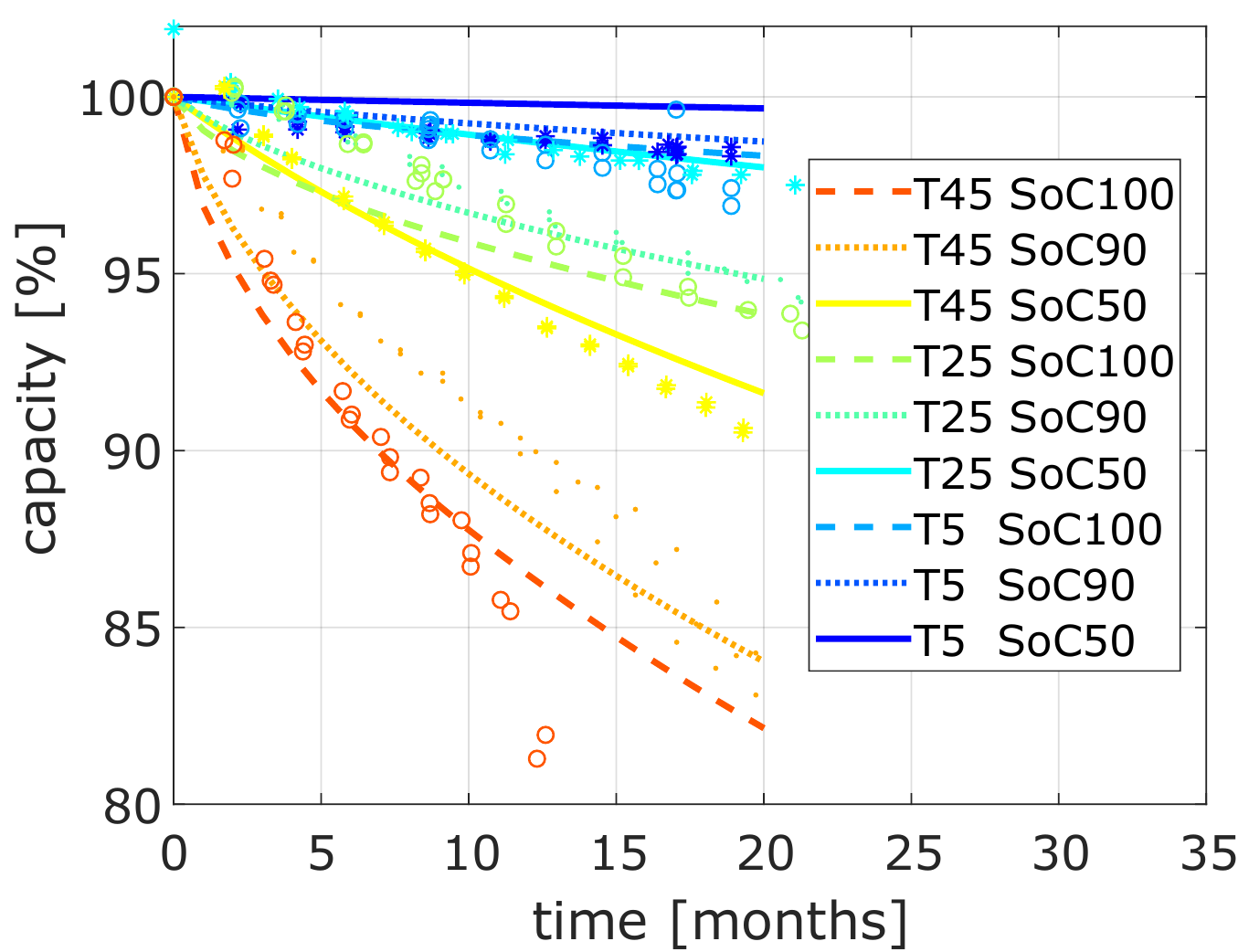}
			\caption{Degradation model parameters were adjusted so that simulations (lines) gave best fit to experimental data (markers), for calendar ageing at various temperatures and SoC levels. Circles and dashed lines are for 100\% SoC; dots and dotted lines are for 90\% SoC; stars and full lines are 50\% SoC.}
			\label{fig:fit_Calendar}
			% location: C:\Users\Jorn Reniers\Desktop\ESO\code\code_desktop\Battery\SPMcell_Mat4Batfit Paper_plotAgeingFit.m
			%TODO got too large font, 12 is correct size [if I have the time and effort to remake this figure]
		\end{figure}

		The parameters of the SPM and degradation models were fitted manually to data from a \SI{16}{Ah} nominal capacity NMC/C Kokam pouch cell \cite{Kokam2016} using the process described in Reniers \cite{reniers2019degradation}. An extensive degradation data set for this cell was collected in the EU Mat4Bat project \cite{EuropeanCommission2016}. The diffusion and rate constants of the SEI growth model (Equation \ref{eqn:SEI}), and the fitting parameter $\beta_1$ of the pore clogging equation (Equation \ref{eqn:SEI_pore}), were adjusted to achieve a best fit between the model simulations and calendar ageing test data, as shown in Fig.\ \ref{fig:fit_Calendar}. When resting at high SoC the SEI growth was mostly diffusion limited, resulting in the typical square-root dependency of capacity on time; when resting at low SoC the kinetics become the limiting factor, resulting in a more linear dependency of capacity on time. During calendar ageing, the concentration is uniform and the stress is zero and therefore the crack-related LAM model has no effect.
	
\begin{figure}
			\centering
			\includegraphics[width=8cm]{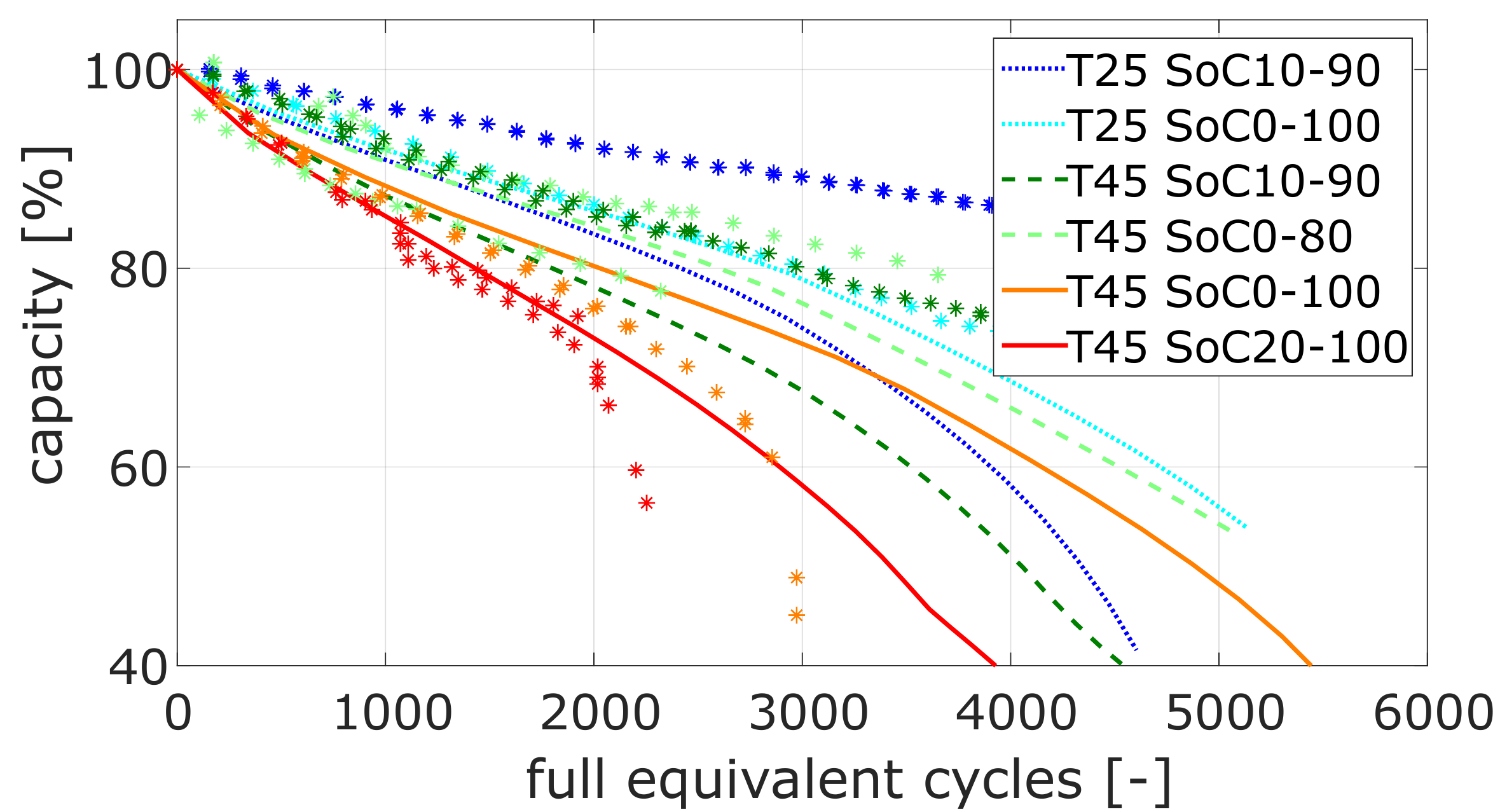}
			\caption{Degradation model parameters were also adjusted to fit simulations (lines) to experimental data (markers), for cycle ageing at various temperatures and SoC windows.}
			\label{fig:fit_Cycle}
		\end{figure}

		When the SEI model parameterised from calendar ageing data was used to simulate cycle ageing, it underestimated the degradation across all test conditions, and especially the degradation from cycling at high SoC levels. However, this is to be expected because other degradation mechanisms are also active during cycling. In our model implementation, the crack-related LAM model increases the degradation that takes place during cycling. The fitting constants for this model, $\beta_2$ and $m$ in Equation \ref{eqn:stress_LAM}, were adjusted such that the combined SEI and crack growth LAM degradation models together approximated the cycle ageing data shown in Fig.\ \ref{fig:fit_Cycle}, where all charging is 1C CCCV and all discharging is 1C CC. The LAM model does not have a strong SoC dependency, therefore not all SoC windows could be fitted accurately. Instead, the parameter fit for the cycling data was focused on maximising the accuracy of the data at \SI{25}{\celsius} in the full SoC window (coloured cyan), because that condition was used throughout the rest of the later studies in this paper. Consequently, degradation for cycling between the more limited 10\% to 90\% SoC window may be slightly overestimated. %
	
\begin{figure}
			\centering
			\includegraphics[width=8cm]{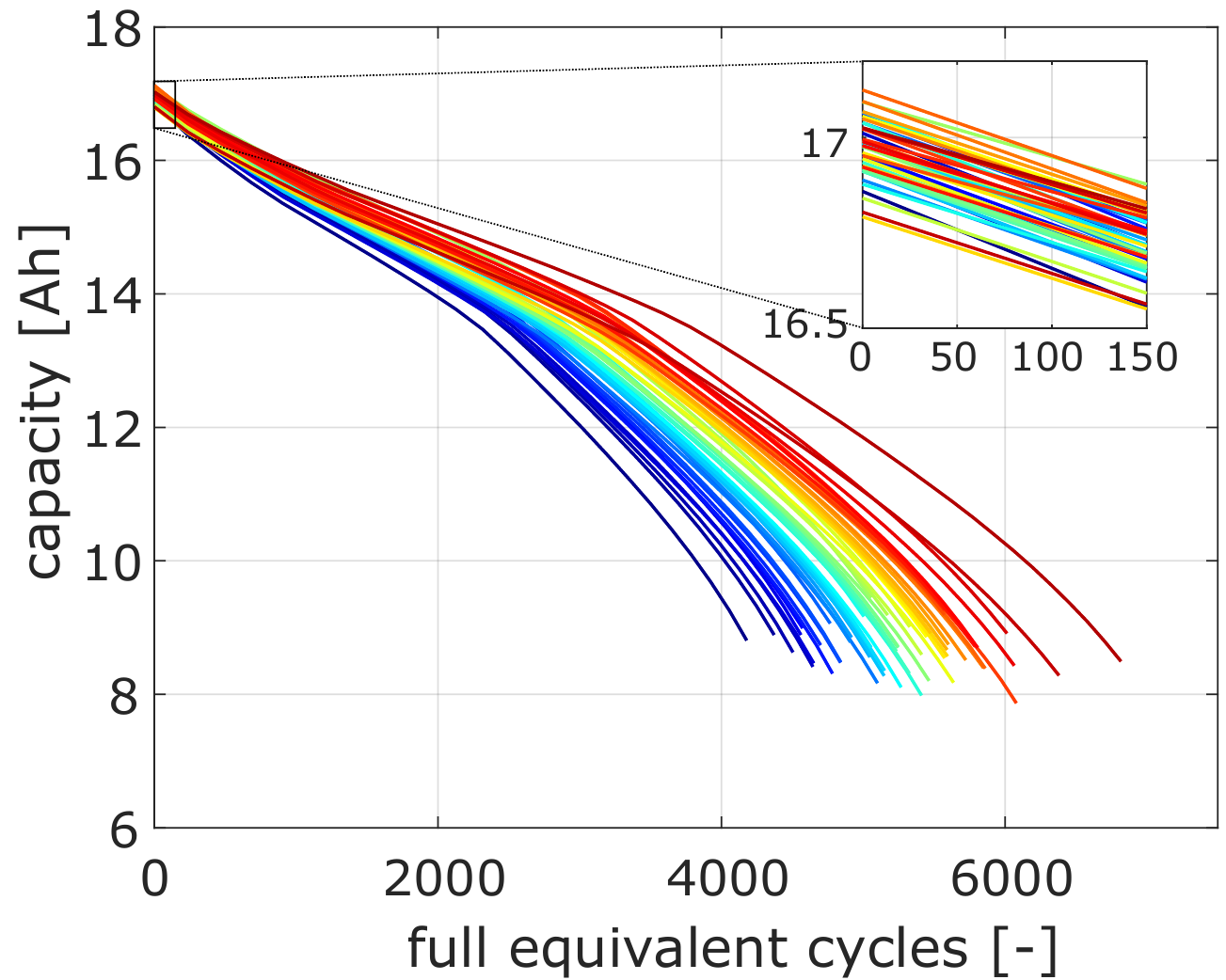}
			\caption{Simulated variability in capacity fade trajectories for 50 cells cycling at 1C CCCV, 25 $^\circ$C over the full SoC window.}
			\label{fig:fit_Variation}
		\end{figure}

		Because a grid-scale battery system consists of thousands of cells, it is important to include cell-to-cell performance variations in the simulation. To model this, firstly, at the beginning of life, cells are expected to have slightly different capacities and resistances. Barreras et al.\ \cite{Barreras2017a} screened a batch of 200 new Kokam cells and found that the measured capacities and resistances were normally-distributed with relative standard deviations of 0.4\% and 2.5\% respectively, so we also assumed normally-distributed initial capacities and resistances with these standard deviations. Secondly, as the cells degrade, differences between them increase, as shown by Harris et al.\ \cite{Harris2017} and Baumh\"ofer et al.\ \cite{Baumhofer2014}. To simulate differing degradation rates, the parameters $D_{\text{sei}}$ and $k_{\text{sei}}$ (Equation \ref{eqn:SEI}) were assumed to be drawn from correlated Gaussian distributions with mean values equal to their respective nominal values, resulting in an overall Gaussian distribution of the SEI side reaction current $i_\text{sei}$. Similarly, the parameter $\beta_2$ (Equation \ref{eqn:stress_LAM}) was assumed to be drawn from a second, uncorrelated Gaussian distribution. %
		In order to produce variations similar to those reported by Harris and Baumh\"ofer, the distributions of these variables were each given a relative standard deviation of 10\%. Under all of these assumptions, Fig.\ \ref{fig:fit_Variation} shows the resulting simulated degradation for 50 cells each individually cycling at 25$^\circ$C over the full SoC window. Note that the four random variables (cell capacity, specific resistance of the electrodes, SEI side reaction current, and LAM rate) were assumed to be independent. 

	\subsection{Electrical coupling}
		\label{elec_coupling}
\begin{figure}
			\centering
			\includegraphics[width=8cm]{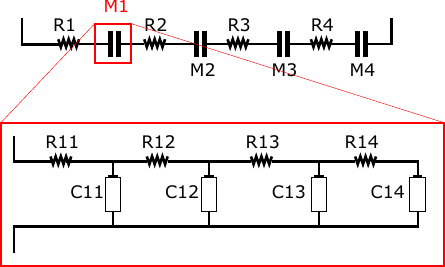}
			\caption{Electrical model for an example 4s4p module; top shows series connections and bottom shows close-up of a parallel-connected block. Resistors represent electrical contact resistances between cells.}
			\label{fig:elec_model}
		\end{figure}

		Parallel and series electrical connections between cells, including contact resistances, were implemented in the Group class, which is hierarchically structured. For parallel-connected groups, it is assumed the terminals are `before' the first cell, i.e.\ to the left of, or upstream of, the first cell, as shown in Fig.\ \ref{fig:elec_model}, an electrical diagram of an exemplary 4s4p module. Here, the top level is a series-connected module and this ensures that the same current runs through series-connected blocks M1, M2, M3, and M4. Within each of the blocks M1-4 is a parallel-connected structure having 4 cells labelled C11-C14, each of which is to be kept at the same voltage. To enforce this constraint mathematically, using block M1 as an example, the current through the $m$th cell within the block is found by solving Equation \ref{eqn:elec_model} (using the convention that positive current is discharging), where $V_{1m}$ is the cell voltage of cell $m$ as given by Equation \ref{eqn:SPM_V}:
		\begin{equation}
		\label{eqn:elec_model}
			V_{11} - R_{11} \sum_{m=1}^{4}I_{1m} = V_{1j} - \left( \sum_{k=1}^{j} \left(  R_{1k} \sum_{m=k}^{4}I_{1m}  \right) \right) \forall j
		\end{equation}
		
		Unfortunately, enforcing these voltage constraints in each parallel group during simulations becomes computationally expensive for large batteries, especially if there are `nested' parallel groups. Iterative methods do not scale well, and even a locally linearised approach proposed by Ashwin et al.\ \cite{Ashwin2017} proved to be infeasible. Instead, to solve this problem a control-theory approach was implemented. Each parallel-connected group includes a proportional-integral (PI) controller for the units (cells or other groups) connected within it, and this controls the currents of parallel-connected unit within the group according to the voltage differences between each parallel-connected unit and the mean voltage of all other cells. For instance, a parallel group of 5 cells contains 5 PI controllers, each keeping the voltage across its respective cell identical to the average of all other cells. If during a discharge the voltage across one cell becomes very large, then the controller of that cell will increase the current through that cell to reduce its voltage, while the controllers of the other cells will reduce the current through their cells to ensure that the total current remains the same. This significantly reduces the overall computational cost since once an equilibrium current split is found, only minor modifications need to be made at each time step. 
		
\begin{figure}
			\centering
			\includegraphics[width=16cm]{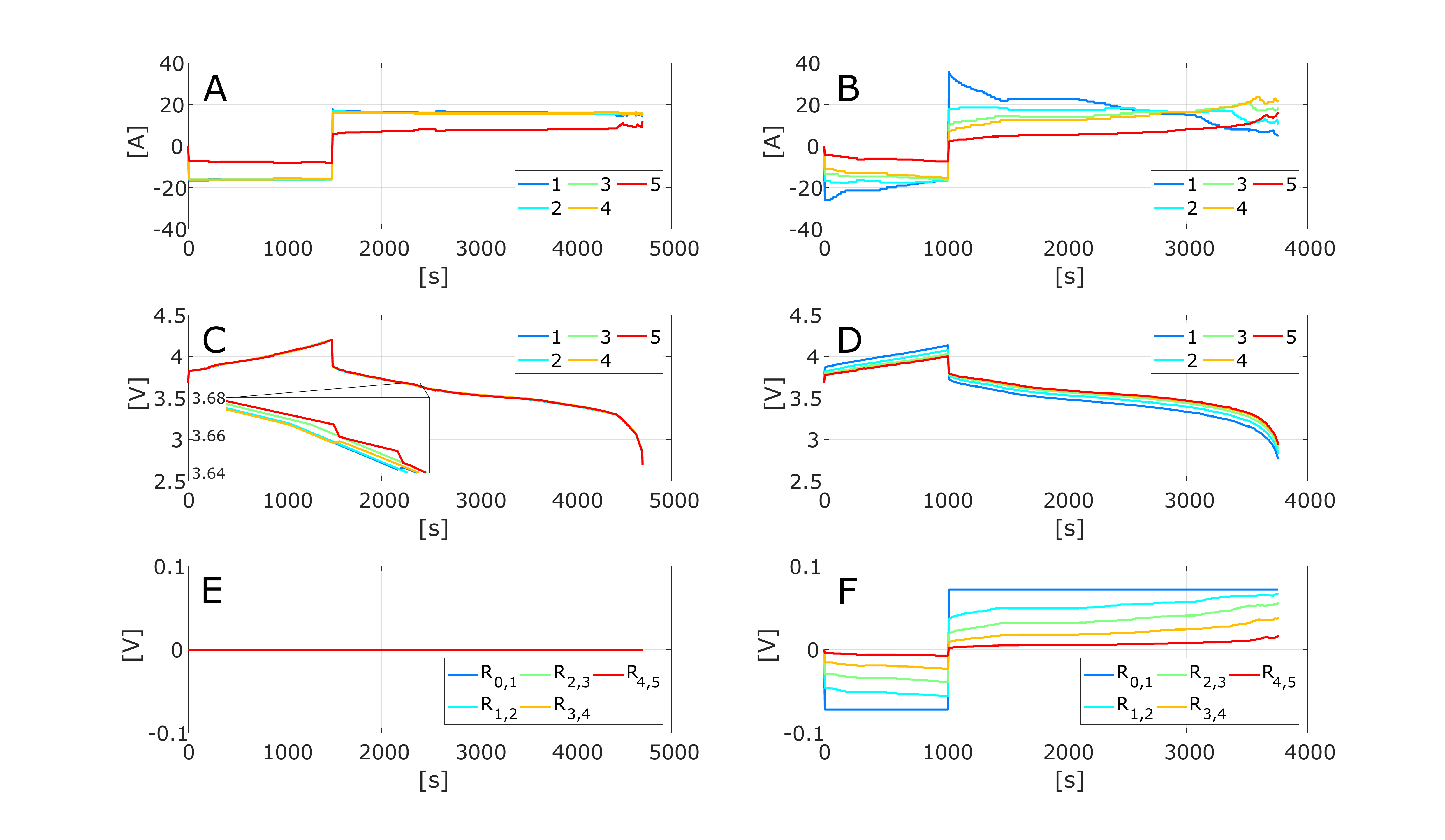}
			\caption{Voltage equalisation in parallel-connected block; fifth cell has half the capacity and double the internal resistance of the others. Left column shows results with no contact resistances, right column shows results with contact resistances of 1 m$\Omega$. A \& B are cell currents; C \& D are cell voltages; E \& F are the voltage drops over the contact resistances between cells.}
			\label{fig:elec_Vequalisation}
		\end{figure}

		Fig.\ \ref{fig:elec_Vequalisation} shows the action of this controller for an example block with 5 parallel-connected cells, 4 having small variations in resistance and capacity between them, and the fifth having half the capacity and double the resistance of the others. % 
		When no electrical contact resistances are included (left column of Fig.\ \ref{fig:elec_Vequalisation}), the cell voltages are within the allowed tolerance of 0.01 \% (Fig.\ \ref{fig:elec_Vequalisation}(c)). The current of the fifth cell is on average half that of other cells, but varies depending on the slope of the open circuit voltage curve and the effect of the cell resistance. When relatively large contact resistances of 1 m$\Omega$ are included (right column of Fig.\ \ref{fig:elec_Vequalisation}),  the cell voltages differ (Fig.\ \ref{fig:elec_Vequalisation}(d)), with the difference between adjacent cells given by the voltage drop over the resistance between them (Fig.\ \ref{fig:elec_Vequalisation}(f)). Since the full block current passes through the first resistor $R_{1,1}$, its voltage drop is constant, unlike the other resistances which only conduct the current of the cells `behind' them. The cell currents shown in Fig.\  \ref{fig:elec_Vequalisation}(b) show that, at the start of discharge, the majority of the current passes through the first cell, as expected in this configuration.

	\subsection{Thermal coupling} \label{thermalCoupling}
	
\begin{figure}
			\centering
			\includegraphics[width=8cm]{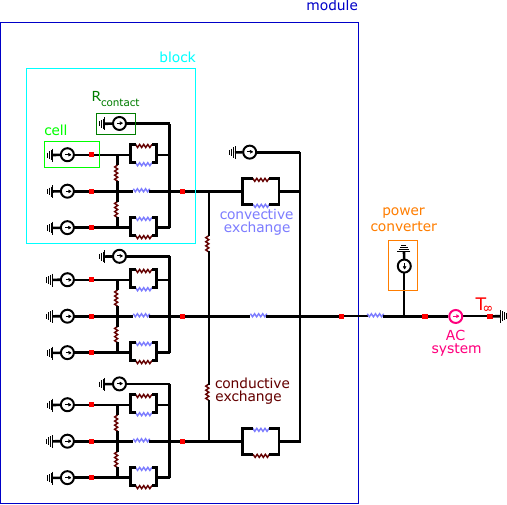}
			\caption{Thermal equivalent circuit model for an example battery consisting of 3 blocks, each with three cells. Red squares indicate points where the thermal equation is solved for the local temperature.}
			\label{fig:thermal_model}
		\end{figure}

	    Temperature variations within and between cells have a significant impact on battery performance. In this work, a lumped thermal model was assumed with individual cells having a single average temperature each---it was not computationally feasible to model temperature variations inside each cell. However, differences in temperatures from cell to cell were modelled using an equivalent circuit network, including heat exchange between adjacent cells and to the cooling system. Fig.\ \ref{fig:thermal_model} shows an example of this for a module made up of three blocks, each consisting of three cells. Individual cell temperature is described by Equation \ref{eqn:SPM_temp} (per cell). Cells exchange heat with  neighbouring cells through the source and sink terms $Q_l=h_l A_{\text{cell}}\left( T - T_l \right)$, which represent conductive heat transfer between adjacent cells, or with the block walls for the first and last cell in a block, or convective heat transfer to the block, i.e.\ the cooling system.
		The thermal system is arranged hierarchically---a Group exchanges heat between its constituent units, has conductive heat transfer with neighbouring Groups, is cooled by convective heat transfer to a higher-level Group, and has additional internal ohmic heating due the contact resistances.
		
\begin{figure}
			\centering
			\includegraphics[width=8cm]{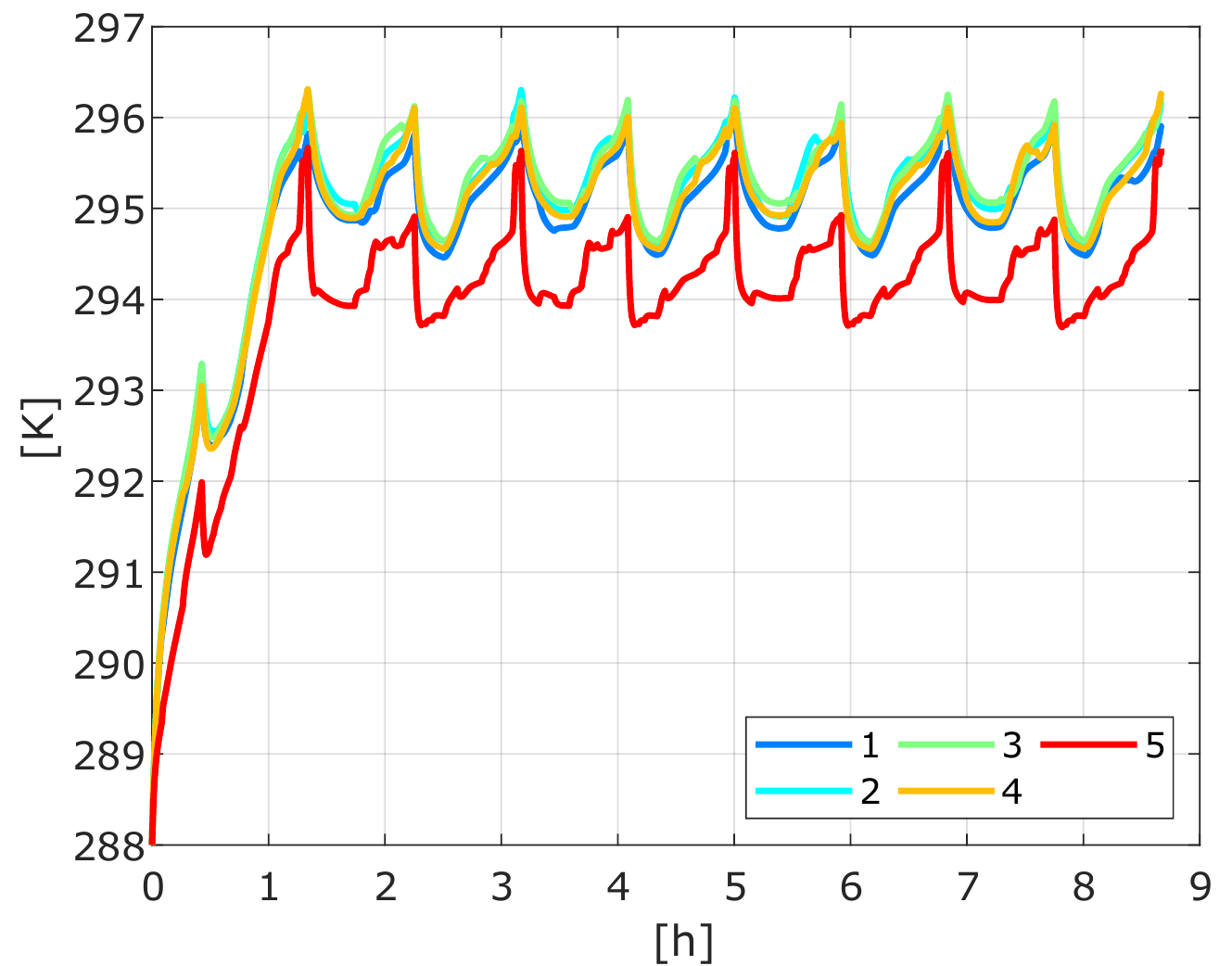}
			\caption{Simulations of cell temperatures (Kelvin) for five parallel-connected cells without contact resistances. Cell number five has half the capacity and double the resistance of the others, which only have small variations in capacity and resistance between them. The block is loaded with five constant current 1C cycles.}
			\label{fig:thermal_T}
		\end{figure}

		The battery container consists of many modules (all thermally in parallel), a power converter, and an air conditioning (AC) system. The battery will heat up from heat exchanged within the battery module and the losses from the power electronic converter, and is cooled by the AC system which removes heat to the environment at $T_{\infty}$. During simulations, the thermal model is resolved `top down' at the same time as the electrical model. %
		Fig.\ \ref{fig:thermal_T} shows simulated cell temperatures during five 1C CC cycles in the same bock as used previously in Fig.\ \ref{fig:elec_Vequalisation}, i.e.\ five parallel-connected cells of which one has half the capacity and double the resistance. The contact resistances were set to zero. The fifth cell has a smaller current, and therefore less heat is generated and this cell has a lower temperature. Cells three to four are slightly hotter than cell one because they are in the middle of the stack and therefore exchange heat conductively with two adjacent hot cells, while cells one and five benefit from some conductive cooling to the wall of the block. Initially the cells heat up until they reach an equilibrium temperature of about \SI{23}{\celsius}. They fluctuate around this equilibrium due to the entropic heating and cooling, and the activity of the cooling system. Details of the cooling system are given in the next section.
		
	\subsection{Ancillary systems} \label{ancillarySystems}
		In addition to the models described above, there are two ancillary systems that are also modelled: the thermal management system and the power electronic converter.
		
		First, the thermal management system. Thermally, Groups can be defined either as `open' or `closed'. An `open' Group simply means there is no barrier between the child units and the parent Group, in other words, the child units can be cooled by the air stream of the parent Group and no additional fan is needed at this level. Alternatively, in the closed case, Groups assume a physical barrier, e.g.\ cells which are enclosed in a metal box, and in this case a fan is needed to cool the child units.
		
		Cooling fans can be actively controlled. If their rotational speed changes, the associated flow rate and air speed changes and consequently the associated convective cooling coefficient will change according to the empirical relation \cite{Toolbox}
		\begin{equation}
		\label{eqn:thermal_h}
			h = 12.12 - 1.16 v + 11.6 \sqrt{v},
		\end{equation}
		where $v$ is the speed of the air, obtained by dividing the flow rate by the cross sectional area of the fan. The energy required to operate each fan is tracked as a source of loss according to 
			\begin{equation}
			\label{eqn:thermal_fanOperate}
			E = \frac{\rho A_{\text{fan}} v^3}{\eta_{\text{fan}}},
		\end{equation}
		where $\rho$ is the density of air, $A_{\text{fan}}$ is the cross-sectional area of the fan, $v$ is the air speed and $\eta_{\text{fan}}$ is the efficiency of the fan. These parameters were assumed as those of a `heavy duty fan' \cite{RSpro2020} with diameter \SI{0.3}{m}, area \SI{0.7}{m^2}, flow rate \SI{65}{m^3/min}, and power consumption \SI{550}{W}. Because lower-level Groups (e.g.\ blocks) would need smaller fans than higher-level groups (e.g.\ racks), the cross sectional areas and flow rates of fans was scaled proportionally to the number of cells that needed to be cooled by each fan, such that the air speeds and convective cooling constants were always in the same range. Schimpe et al.\ \cite{Schimpe2018c} reported that a rack of \SI{24}{kWh} of batteries needed a fan consuming about \SI{80}{W}, therefore a \SI{550}{W} fan could cool about 2750 cells of \SI{60}{Wh} each, or alternatively, per cell a fan with cross-sectional area of \SI{2.5e-5}{m^2} and flow rate of \SI{5e-4}{m^3/s} is needed.
		
		At the container level, two operational modes are possible for cooling. In the first case, if the external environmental temperature is sufficiently low, a container may be cooled directly by ingesting cold outside air and venting warmer air. In this case, the fan was sized exactly as already described, i.e.\ its surface area was scaled according to the number of cells in the container, and the power requirement for cooling was calculated by considering the fan power required, which is based on the difference in thermal energy between the cold and warm air flows, i.e.\ $\dot{m}c_p\left( T_\text{hot}-T_\text{cold} \right)$ where $\dot{m}$ is the air mass flow rate and $c_p$ the heat capacity of air. In the second case, where the external temperature is too hot for direct cooling, an active air conditioning chiller unit must be used. In this case, the operating power is the product of the coefficient of performance, assumed to be 3 \cite{Schimpe2018c}, and the required cooling power. 
		
		The second ancillary system to be modelled is the power electronic converter which transforms the variable DC voltage from the batteries to a fixed AC voltage at  the grid connection. The `average model' of Patsios et al.\ \cite{Patsios2016} was assumed---this includes a two stage converter, which has a DC/DC step to transform the variable DC voltage to a fixed value, and a DC/AC step to enable grid connection. The conduction losses for each stage are given by equation \ref{eqn:converter_conduction}, where $I$ is the current at the stage, respectively the battery and intermediate DC bus currents, $V_{sc}$ is the voltage drop over the semiconductor switches, and $D$ is the modulation ratio: 
		\begin{equation}
			\label{eqn:converter_conduction} 
			P_{\text{cond}} = I V_{sc} D
		\end{equation}
		The switching losses for each stage are given by (\ref{eqn:converter_switch}), where $f$ is the switching frequency, $E_{\text{on}}$ and $E_{\text{off}}$ are the switch-on and switch-off losses:
		\begin{equation}
		\label{eqn:converter_switch} 
			P_{\text{sw}} = f \left( E_{\text{on}} + E_{\text{off}} \right)
		\end{equation}
		The third set of power converter losses occur in the passive elements, i.e.\ the DC/DC converter filter, DC bus capacitor, and DC/AC filter, and were all calculated according to the methods described in Patsios et al.\ \cite{Patsios2016}. In all cases, parameter values from Patsios et al.\ \cite{Patsios2016} were assumed, and where none were provided, those of Schimpe et al.\ \cite{Schimpe2018c} were assumed. Values were rescaled to obtain a converter of the appropriate power rating.
		
\begin{figure}
			\centering
			\includegraphics[width=8cm]{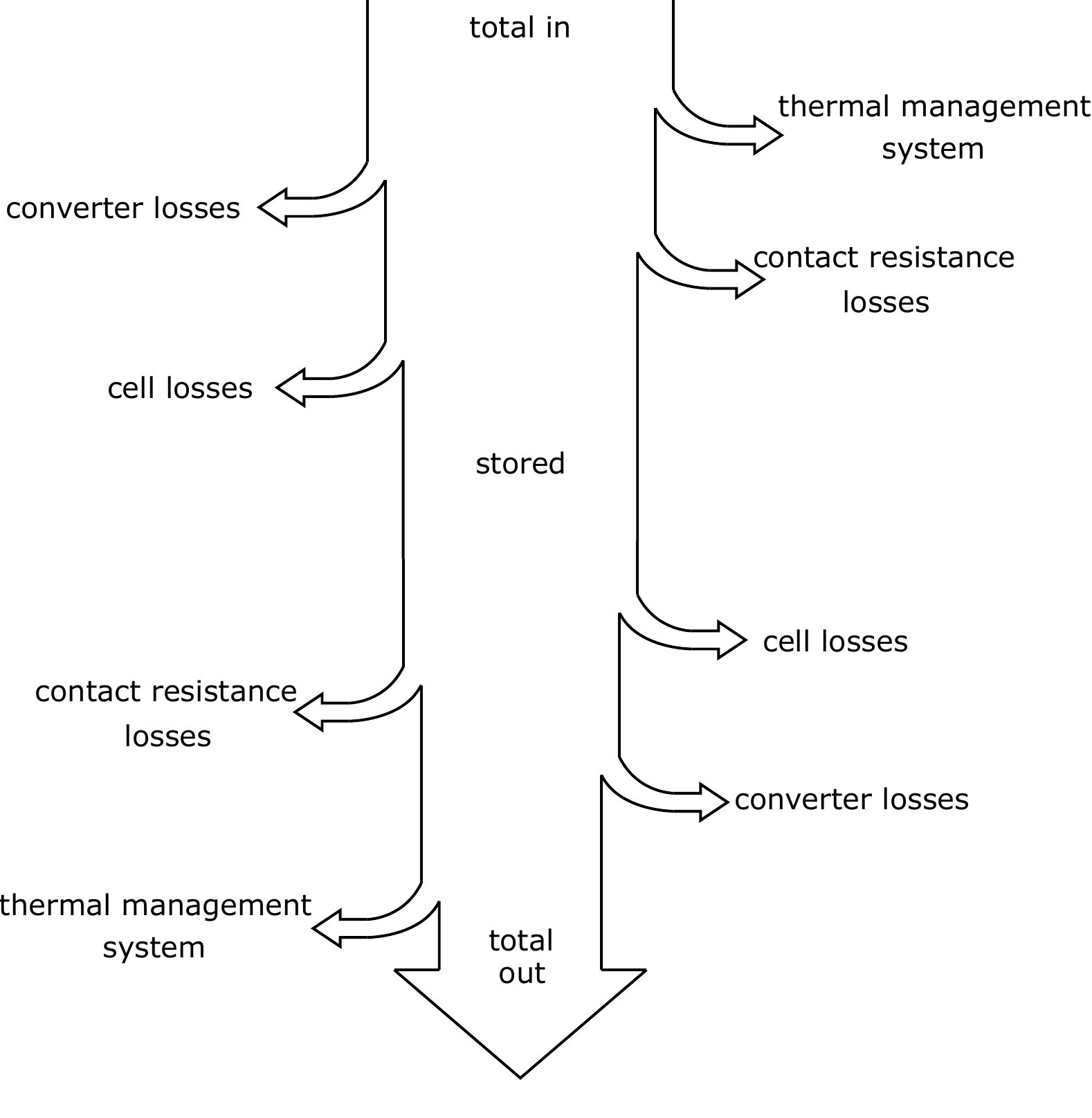}
			\caption{Sankey diagram of the battery energy flow and losses during a round-trip cycle.}
			\label{fig:sankey_diagram}
		\end{figure}

		Summarising all the energy losses in system that occur in a `round trip' (i.e.\ charging followed by discharging), a Sankey diagram of the battery is given in Fig.\ \ref{fig:sankey_diagram}. %
		During charging, AC power is required for the operation of the cooling system, and the remaining power is converted to DC with associated converter losses. The DC power is distributed to the cells, with ohmic losses in the contact resistances and in the cells themselves, and the remaining energy is stored in the cells. During discharging, this process is reversed and the remaining energy is sent back to the grid. The round-trip efficiency is calculated at the grid-interface, i.e.\ total energy out from a full discharge as a fraction of total energy in during a full charge. Similarly, losses are always expressed relative to the total charging energy (`total in') over the same time period. 
		
\section{Results and discussion}

	We now present and discuss the results of various simulation studies, focusing on degradation and round-trip efficiency, beginning with a very simple approach and then investigating the impact of adding more complexity and heterogeneity step-by-step. The overall battery architecture comprises cells connected into 20s7p modules, racks made of 15 series-connected modules, and finally a battery container consisting of 9 parallel-connected racks, giving 18900 cells in total. In the first three sections below (\ref{sec:contactR}--\ref{sec:temp}), when the effect of various submodels was analysed, the battery system was assumed to cycle continuously at 1C constant current, and both charging and discharging were stopped as soon as at least one cell inside the container reached its voltage limit. However, in section \ref{section:thermal_management_control} the battery was loaded with a more realistic current profile to analyse the performance of the full model in a realistic scenario.
	
	\subsection{Electrical contact resistances (isothermal case)} \label{sec:contactR}
	First, the effect of the pack electrical model is analysed, assuming isothermal behaviour, i.e.\ with  thermal models for the cell and overall battery switched off. A simplest case baseline was established by simulating a single cell and multiplying its current and voltage respectively by the number of cells in parallel and series within the pack, plus accounting for the power converter losses. This is referred to as the `1 cell'-model. In addition to this, to investigate the impact of electrical contact resistance variations, a separate simulation was established that modelled every cell individually, but with identical cell parameters and different electrical contact resistances between cells, modules, and packs. 
	
\begin{figure}
		\centering
		\includegraphics[width=16cm]{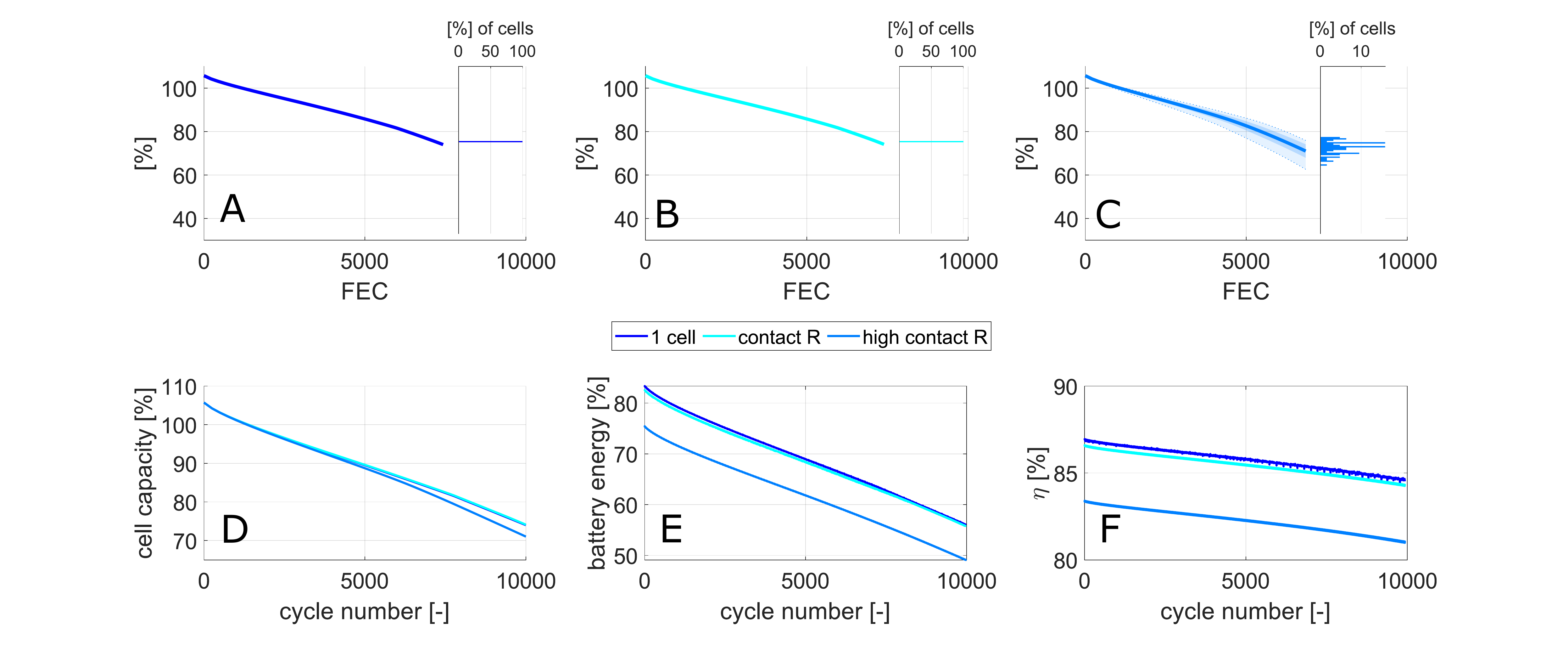}
		\caption{Impact of electrical contact resistances on system performance. Results for three scenarios: a single cell model scaled-up (`1 cell'); identical cells connected via contact resistances (`contact R'); identical cells connected via ten times larger contact resistances (`high contact R'). A, B and C show the evolution of the cell charge capacities for the three models (solid line is mean, dark shaded area is mean plus or minus one standard deviation, light shaded area and dotted line are lowest and highest cell capacities) and histogram of final capacity distribution; D compares mean cell capacities of the three models; E shows 1C discharge energy; F shows round-trip efficiency.}
		\label{fig:complex_elec_model}
	\end{figure}

	The results of this electrical study are compared in Fig.\ \ref{fig:complex_elec_model}. The `1 cell' model fades to 80\% capacity in around 6000 cycles. The `contact R' simulation had electrical contact resistances set at realistic values as reported in the literature by Schimpe et al.\ \cite{Schimpe2018c}, who investigated the resistances within a \SI{192}{kWh} battery, measuring values of \SI{0.0075}{\milli\ohm} for connections of cells in blocks, and blocks in modules, and \SI{0.25}{\milli\ohm} for higher-level connections (modules-to-racks, and racks to main DC bus in the battery container). A second case was simulated with ten times higher contact resistances throughout (`high contact R'). As an aside, these higher values are still below the values used by some studies investigating the effect of contact resistances, e.g.\ Liu et al.\ \cite{Liu2019} used \SI{10}{\milli\ohm}, Rumpf et al.\ \cite{Rumpf2018} used \SI{0.9}{\milli\ohm} to connect cells in blocks, and Schindler et al.\ \cite{Schindler2020} used values of 15-\SI{62}{\milli\ohm}. However in our view, these high values are not realistic in stationary battery applications due to the high currents and large losses involved---the battery usable energy would be reduced to unrealistically low values even without accounting for the additional operating energy of the thermal management system.
	
	Fig.\ \ref{fig:complex_elec_model}.A--C show the evolution and distribution of the cell capacities versus full equivalent cycles (FEC), where capacity is defined as the charge which can be accepted between the voltage limits of the cell during a CCCV charge, divided by the nominal capacity of the cell. As can be seen, there is almost no difference between simulating one cell and simulating all cells but with small contact resistances. On the other hand, large contact resistances increase the overall degradation rate and also increase the cell-to-cell variability in later life due to the inhomogeneous current distribution.
	
	Fig.\ \ref{fig:complex_elec_model}.D compares the mean cell capacity of all three models versus cycle number, and it can be seen that the mean cell capacity of the high-contact-resistance model is about 3 \%-pts lower than the other two models after 10000 cycles. %
	The relative usable energy of the entire battery, shown on Fig.\ \ref{fig:complex_elec_model}.E, is the energy which can be discharged at constant 1C current after subtracting all losses, i.e.\ the `total out' energy shown in Fig.\ \ref{fig:sankey_diagram}, divided by the total energy capacity of the battery, the latter being the product of the nominal cell capacity, the nominal cell voltage and the number of cells. Due to losses and diffusion limitations, even at the start of life the usable energy is well below 100 \% of the total energy available. Again, using  realistic values for contact resistances has almost no impact on this result compared to the `1 cell' model. However, in the case with high contact resistances, the energy is about 7 \%-pts lower over the entire lifetime, mostly due to the high voltage drops over the contact resistances---less energy can be added to the battery during charging  before the voltage limits are reached due to the voltage drops over the resistances, and during the subsequent discharge, more energy is lost due to ohmic heating in the resistances. The increased degradation and cell-to-cell variations have little effect on the usable energy. %
	Finally, the round trip efficiency shown on Fig.\ \ref{fig:complex_elec_model}.F is the ratio of the discharged to the charged energy during a 1C CC cycle, both measured at the interface to the outside world as explained in section \ref{ancillarySystems}. The large contact resistances decrease efficiency by about 3 \%-pts over the entire lifetime.
	
	\subsection{Cell-to-cell variations (isothermal case)} \label{sec:cell2cell}
	The second simulation study explores the impact of cell-to-cell variations on ageing. Here, the base case (`identical') is the scenario where every cell is simulated individually, with identical model parameters, and with realistic contact resistances (`contact R', Fig.\ \ref{fig:complex_elec_model}). The other scenarios add variations in the initial values of three model parameters, namely cell DC resistance ($r_{\text{dc},n}$ and $r_{\text{dc},p}$ from (\ref{eqn:SPM_Rdc})), cell capacity (i.e.\ surface area of the electrodes, $A_n$ and $A_p$ from (\ref{eqn:SPM_BCsurface})), and the rate of degradation ($k_{\text{sei}}$ and $D_{\text{sei}}$ from (\ref{eqn:SEI}), $\beta_2$ and $m$ from (\ref{eqn:stress_LAM})). As explained in section \ref{sec:parameterisation}, the initial distribution of cell resistance, capacity and degradation rate was assumed to have a standard deviation of 2.5\%, 0.4\%, and 10\% respectively, and each distribution is independent of the others.
	
\begin{figure}[t]
		\centering
		\includegraphics[width=16cm]{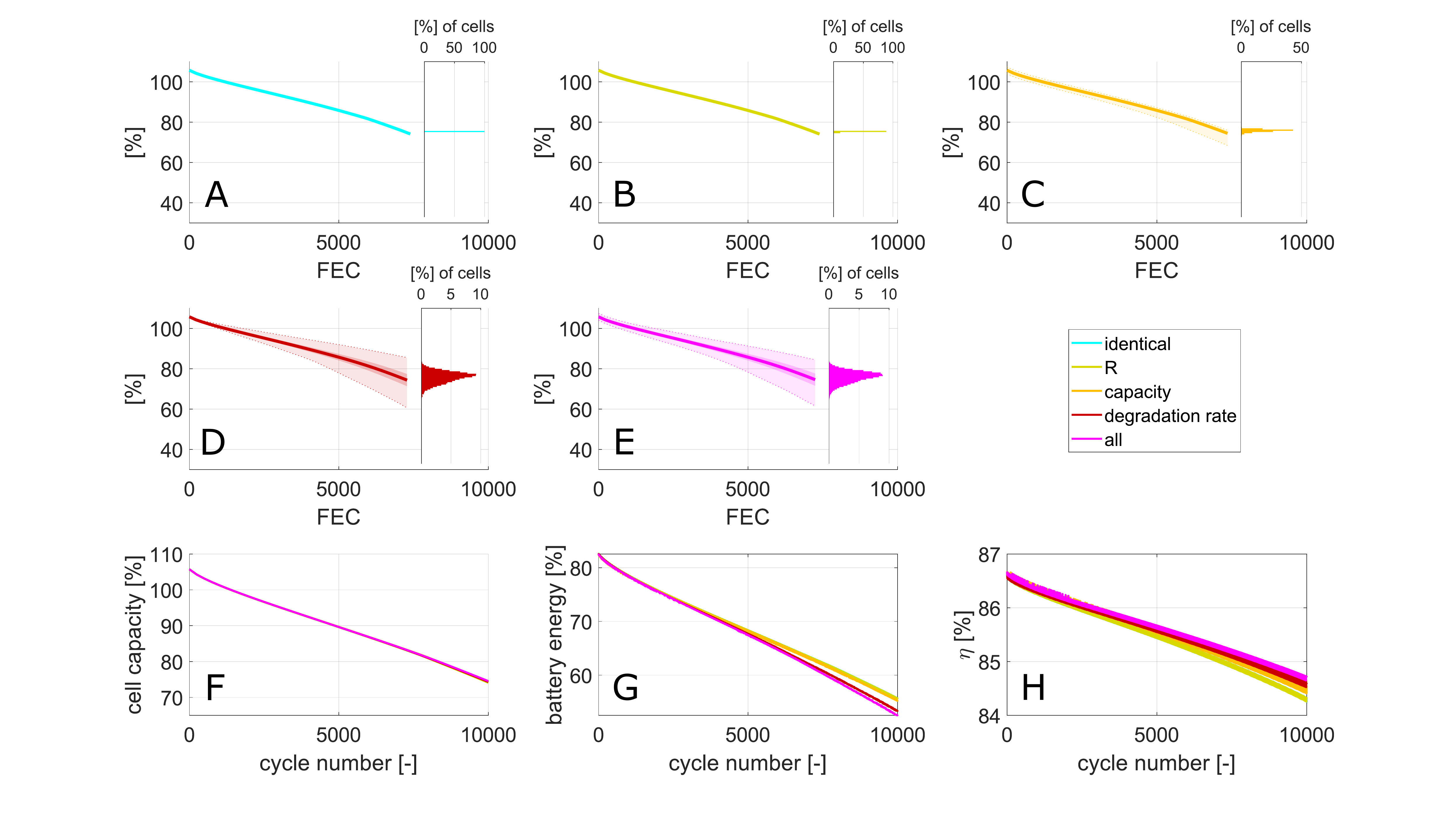}
		\caption{Impact of cell-to-cell variations on system performance, including nominal contact resistances. Scenarios: all identical cells (`identical'); with distribution in cell resistance (`R'); with distribution in cell capacity (`capacity'), with distribution in cell degradation rate (`degradation rate'); with distribution in all three categories (`all'). A--E show the evolution of charge capacities for all five models (solid line is mean, dark shaded area is mean plus or minus one standard deviation, light shaded area and dotted line are lowest and highest cell capacities) and histogram of the final capacity distribution; F compares mean cell capacities; G shows 1C discharge energy; H shows round-trip efficiency.}
		\label{fig:complex_variation}
	\end{figure}

	In the results of this study, Fig.\ \ref{fig:complex_variation}.A is identical to Fig.\ \ref{fig:complex_elec_model}.B and shows the behaviour assuming all cells are identical and the contact resistances between them are realistic. Fig.\ \ref{fig:complex_variation}.B to Fig.\ \ref{fig:complex_variation}.E show the evolution and spread of the cell charge capacities versus full equivalent cycles for the scenarios where there are, respectively, cell-to-cell variations in internal resistance, capacity, degradation rate, and all of these. The spread in initial resistance has almost no impact on long term performance, and the spread in capacity has only very limited impact. However, the spread in degradation rate leads to a significant increase in cell-to-cell variations, especially later in life. This confirms the findings of Zilberman et al.\ \cite{Zilberman2020}, namely that the spread in degradation rate is by far the most influential factor for pack lifetime. The terminal capacity is still normally distributed because the (almost linear) SEI growth dominates the degradation for the simulations considered here---the LAM model only starts to dominate behaviour significantly below 80\% capacity. The LAM-degradation model has a pronounced `knee-point' after which degradation rapidly increases, which would alter the shape of the distribution since cells `beyond' the knee would degrade much more rapidly, giving rise to a long tail of cells with low capacities. When the spread in all three parameters are considered, the results barely change because the spread in degradation rate dominates the other two. %
	
	Although the mean cell capacity is identical in all cases (Fig.\ \ref{fig:complex_variation}.F), the system usable energy (Fig.\ \ref{fig:complex_variation}.G) differs between scenarios because each series-connected rack is limited by its weakest block---the usable energy is a function of both the mean capacity and the spread in the degradation rates. Compared to having all identical cells, or only a spread in resistance (which both give the same results), the spread in cell capacity reduces usable energy by 0.5 \%-pts after 10,000 cycles, the spread in degradation rates leads to a 2.5 \%-pts reduction, and the spread in all three parameters to a 3 \%-pts reduction. The differences in the overall efficiency (Fig.\ \ref{fig:complex_variation}.H) are small (0.5 \%-pts maximum difference), but in the opposite order, i.e.\ the case with spread in all three variables has the highest round-trip efficiency compared to having all-identical cells which has the lowest round-trip efficiency. This is due cells with lower degradation rate and resistance tending to pass more of the current, reducing the overall system losses.
	
	\subsection{Temperature variations} \label{sec:temp}
	Our third simulation study explores the degradation impact of non-isothermal behaviour. The base-case is a model with realistic electrical contact resistances plus a spread in initial cell resistance, capacity, and degradation rate. The simplest thermal model is one that accounts for individual cell temperatures---Equations (\ref{eqn:SPM_temp}) and (\ref{eqn:SPM_arrhenius})---but ignores inter-cell coupling and the thermal management system (section \ref{thermalCoupling}). Instead, it is assumed each cell is cooled by convection with air at the environmental temperature. This base-case model is referred to as the `individual cell'-thermal model. As an alternative, a full-scale model was also implemented, adding both the thermal coupling between adjacent cells/units, and the thermal management system to transport heat from the cells to the environment. This is referred to as the `coupled + cooling system'-model.
	
\begin{figure}
		\centering
		\includegraphics[width=16cm]{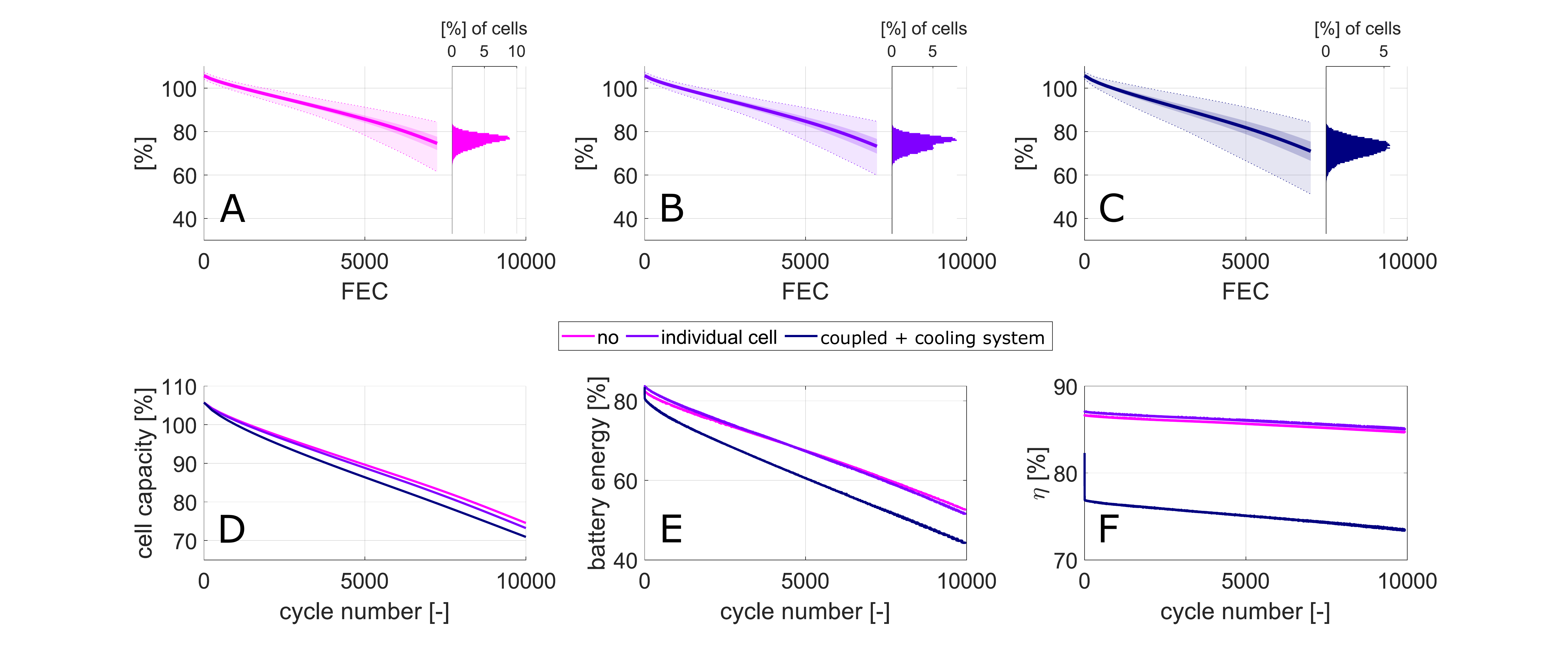}
		\caption{Impact of thermal management system on system performance, including contact resistances and cell-to-cell variations in resistance, capacity and degradation rate. Scenarios: no thermal model (`no'); individual cell thermal models (`individual cell'); fully coupled thermal model and cooling system (`coupled + cooling system'). A--C show evolution of cell charge capacities (solid line is mean, dark shaded area is mean plus or minus one standard deviation, light shaded area and dotted line are lowest and highest cell capacities) and histogram of final capacity distribution; D compares mean cell capacities; E shows 1C discharge energy; F shows round-trip efficiency.}
		\label{fig:complex_thermal}
	\end{figure}

	The results of the thermal study are shown in Fig.\  \ref{fig:complex_thermal}, where the base-case, sub-figure A, is identical to Fig.\ \ref{fig:complex_variation}.E. The evolution of the distribution of the cell capacities changes little when the individual thermal models are included, as shown in Fig.\ \ref{fig:complex_thermal}.B. When the fully coupled thermal model is used (Fig.\ \ref{fig:complex_thermal}.C), cells will be at higher temperatures because heat needs to be evacuated through the entire battery. This leads to both higher mean degradation and increased cell-to-cell variations during ageing; the standard deviation in capacity after 10000 cycles (about 7000 full equivalent cycles) increased from 3.0\% to 4.4\%, while the minimum cell capacity decreased from 61.5\% to 51.2\%. % this is base case, i.e. subplot A, compared to coupled, i.e. subplot C
	After 10000 cycles, the mean capacity (Fig.\ \ref{fig:complex_thermal}.D) decreased from a base-case (isothermal) result of 74.6\% to 73.2\% when using individual cell thermal models, and 70.9\% for the fully coupled model.
	
	The change in usable energy was more dramatically different between the three models, as shown in Fig.\  \ref{fig:complex_thermal}.E. Part of this is due to the decreased mean cell capacity and increased cell-to-cell spread, but the majority of the difference is due to the energy required to run the cooling system in the fully cooled case. In the case of the coupled thermal model, about 5\% of the battery discharge energy is required for cooling power. This is also clearly visible in the round-trip efficiency plot (Fig.\  \ref{fig:complex_thermal}.F), where it can be seen that operation of the cooling system reduces round-trip efficiency by about 10 \%-pts (5\% losses during both charge and discharge respectively).
		
	\subsection{Control of thermal management system} \label{section:thermal_management_control}
	Given the critical impact of the thermal management system on both lifetime and energy efficiency, we now consider a simulation study of different thermal control strategies for grid storage. As a reminder, the cooling system consists of an AC unit which cools the battery container by exchanging heat with the environment, and various fans inside the battery that distribute cool air to the cells.
	
	The outside temperature is assumed to be cold enough (\SI{15}{\celsius}) that the AC system switches to a mode where it can use a large fan to suck in cold outside air, and evacuate hot air to the outside, without active chilling via a heat pump. The latter can easily be simulated, and results in an efficiency drop overall, but for simplicity was ignored here. A second large fan circulates air within the battery compartment, and every module (containing 20s7p cells) also has a small fan which takes air from the battery compartment and blows it over the cells. All fans can be controlled individually, and  in this simulation study we compare five different control strategies. In the following descriptions, the `local temperature' refers to the temperature at the fan or at the AC unit, while the `hot-spot temperature' refers to the hottest element `behind' the fan, i.e.\ the hottest cell in the module (for the module fans) or the hottest cell in the battery compartment (for the fan in the battery compartment and the AC unit). The five control strategies considered are as follows:
	\begin{enumerate}
		\item `Always on': All fans are continuously on at full power, and the AC system is on at full power except when the local temperature goes below \SI{20}{\celsius}, in which case it is switched off.		
		\item `Local temperature on/off': Fans operate at full power when the temperature at the fan exceeds \SI{35}{\celsius} and stay on until it goes below \SI{25}{\celsius}; the AC system starts cooling at full power when the local temperature exceeds \SI{25}{\celsius} and switches off when the battery has cooled down to \SI{20}{\celsius}.
		\item `Hot-spot on/off': Fans start operating at full power when the hot-spot temperature behind the fan exceeds \SI{35}{\celsius} and switch off when it goes below \SI{25}{\celsius}; the AC system cools at full power if the total hot-spot temperature exceeds \SI{30}{\celsius}, and switches off when the local temperature has gone below \SI{20}{\celsius}. 
		\item `Proportional to local temperature': Fans operate with power proportional to how much the local temperature $T_{\text{local}}$ exceeds \SI{25}{\celsius}, with a proportional gain such that they will be at full power if the local temperature exceeds \SI{35}{\celsius} as per Equation \ref{eqn:control4_fan}. Similarly, the AC unit's cooling power is proportional to how far the local temperature exceeds \SI{20}{\celsius} and will be at full power when it exceeds \SI{25}{\celsius}, as per Equation \ref{eqn:control4_ac}, 
		\vspace{-0.3cm} % THIS IS A HACK!
		\begin{equation}
			\label{eqn:control4_fan}
			\frac{p_{\text{fan}}}{p_{\text{fan, nom}}} = \text{min} \left(\text{max}\left( \frac{T_{\text{local}} - 25}{35 - 25}, 0 \right), 1\right)
		\end{equation}
		\begin{equation}
			\label{eqn:control4_ac}
			\frac{p_{\text{AC}}}{p_{\text{AC, nom}}} = \text{min} \left(\text{max}\left( \frac{T_{\text{local}} - 20}{25 - 20}, 0 \right), 1\right).
		\end{equation}
		\item `Proportional to hot-spot': Fans operate with power proportional to how far the hot-spot temperature $T_{\text{hot}}$ exceeds \SI{25}{\celsius}, and will be at full power when the hot-spot temperature exceeds \SI{35}{\celsius} as given by Equation \ref{eqn:control5_fan}. Similarly, the AC unit's cooling power is proportional to how far the hot-spot temperature exceeds \SI{25}{\celsius} and will be at full power when it exceeds \SI{30}{\celsius}, but also switches off if the local temperature goes below \SI{20}{\celsius} as given by Equation \ref{eqn:control5_ac},
		\vspace{-0.3cm} % THIS IS A HACK!
		\begin{equation}
			\label{eqn:control5_fan}
			\frac{p_{\text{fan}}}{p_{\text{fan, nom}}} = \text{min} \left(\text{max}\left( \frac{T_{\text{hot}} - 25}{35 - 25}, 0 \right), 1\right)
		\end{equation}
		\begin{equation}
			\label{eqn:control5_ac}
			\frac{p_{\text{AC}}}{p_{\text{AC, nom}}} = \left( T_{\text{local}} > 20 \right) \left( \text{min} \left(\text{max}\left( \frac{T_{\text{hot}} - 25}{30 - 25}, 0 \right), 1\right) \right).
		\end{equation}
		
	\end{enumerate}

\begin{figure}
		\centering
		\includegraphics[width=16cm]{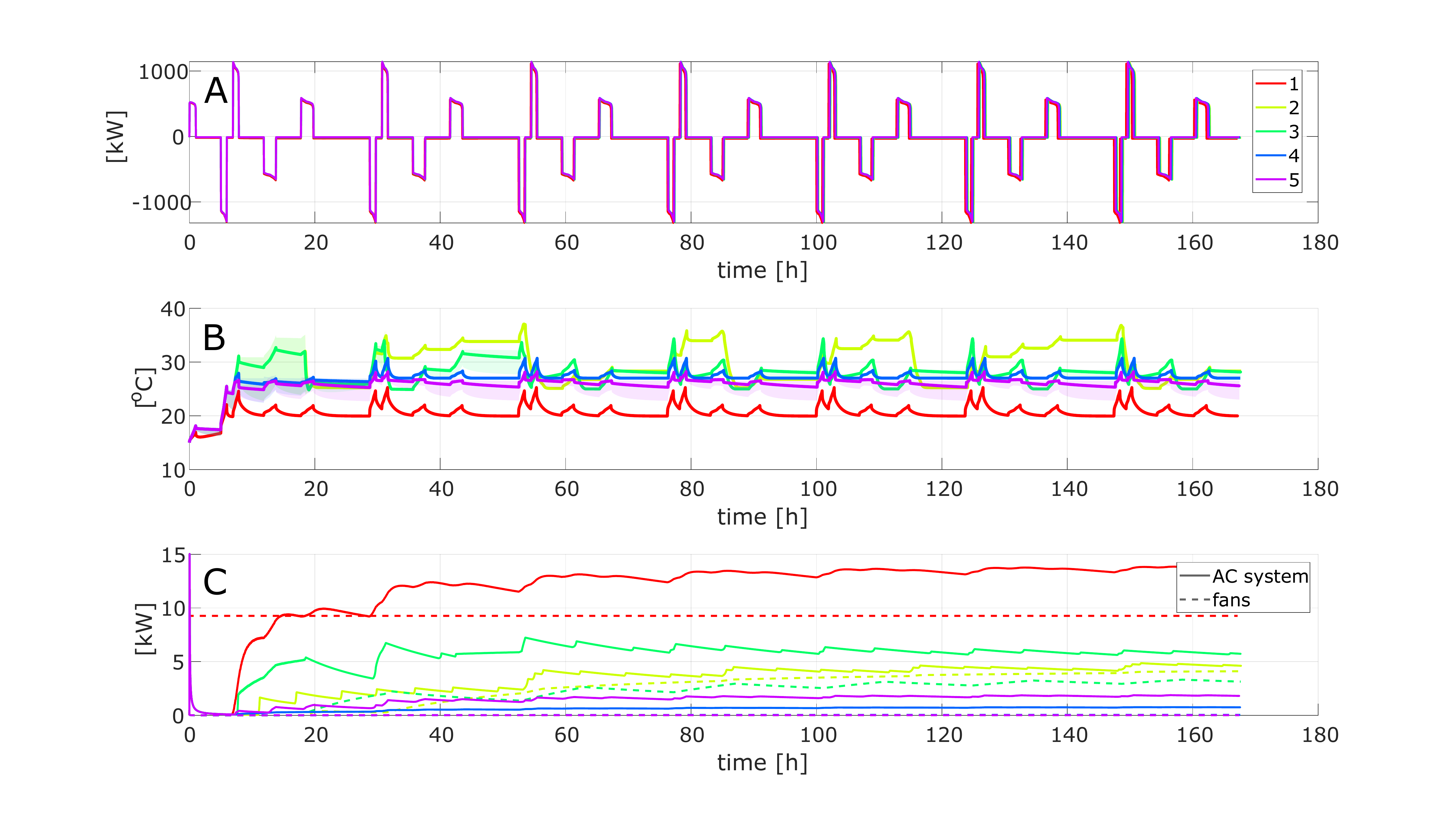}
		\caption{Comparison of five different thermal management approaches---simulations of first week of battery operation. A is battery power output at the grid-interface; B is cell temperatures, with mean indicated by lines and shaded areas giving range between minimum and maximum; C is cooling system operating power separated between AC unit and fans.}
		\label{fig:control_time}
	\end{figure}

	For this study, batteries were cycled with a load profile consisting of two cycles a day; from midnight, the system rested for 4 hours, then charged at 1C, rested for 1 hour, discharged at 1C, rested for 4 hours, charged at 0.5C, rested for 4 hours, discharged at 0.5C, rested for 5 hours. Once a week, all cells were brought to the same voltage to ensure the system remained balanced. Fig.\ \ref{fig:control_time} shows the simulation results for the first week of operation, for all five thermal management approaches. %
	
	At the start of the simulation, the entire battery is at \SI{15}{\celsius} such that no cooling is needed in the first few hours. Method 1 (`always on') requires the most energy to operate the cooling system, both for the AC unit and the fans (Fig.\ \ref{fig:control_time}.C), but it results in the lowest mean cell temperature (Fig.\ \ref{fig:control_time}.B). Method 2 (`local on/off') results in long periods of high cell temperatures when the cells themselves are already warm, but the rest of the battery is still heating up. The AC cooling system is only triggered when the entire battery has heated up to \SI{25}{\celsius}, at which point the cells can reach temperatures above \SI{35}{\celsius}. Method 3 (`hot-spot on/off') results in shorter high-temperature spikes because the cooling system is triggered as soon as at least one cell heats up. This leads to a wider range of cell temperatures, with about \SI{2}{\celsius} difference between the coldest and hottest cell, because cells and modules further from the hot-spot are colder, but receive the same cooling as the hottest cell. Methods 4 and 5 (`proportional to local' and `proportional to hot-spot' respectively) require much less cooling power because they can remove heat more efficiently by running fans at lower speeds---the fan power scales with the air speed cubed (Equation \ref{eqn:thermal_fanOperate}) while the convection  constant scales approximately linearly with air speed. Method 5 results in slightly lower cell temperatures compared to method 4 for the same reason as method 3 compared to method 2. Method 5 also results in the most inhomogeneous temperature distribution, with about \SI{4}{\celsius} difference between the coldest and hottest cell.
	
\begin{figure}
		\centering
		\includegraphics[width=8cm]{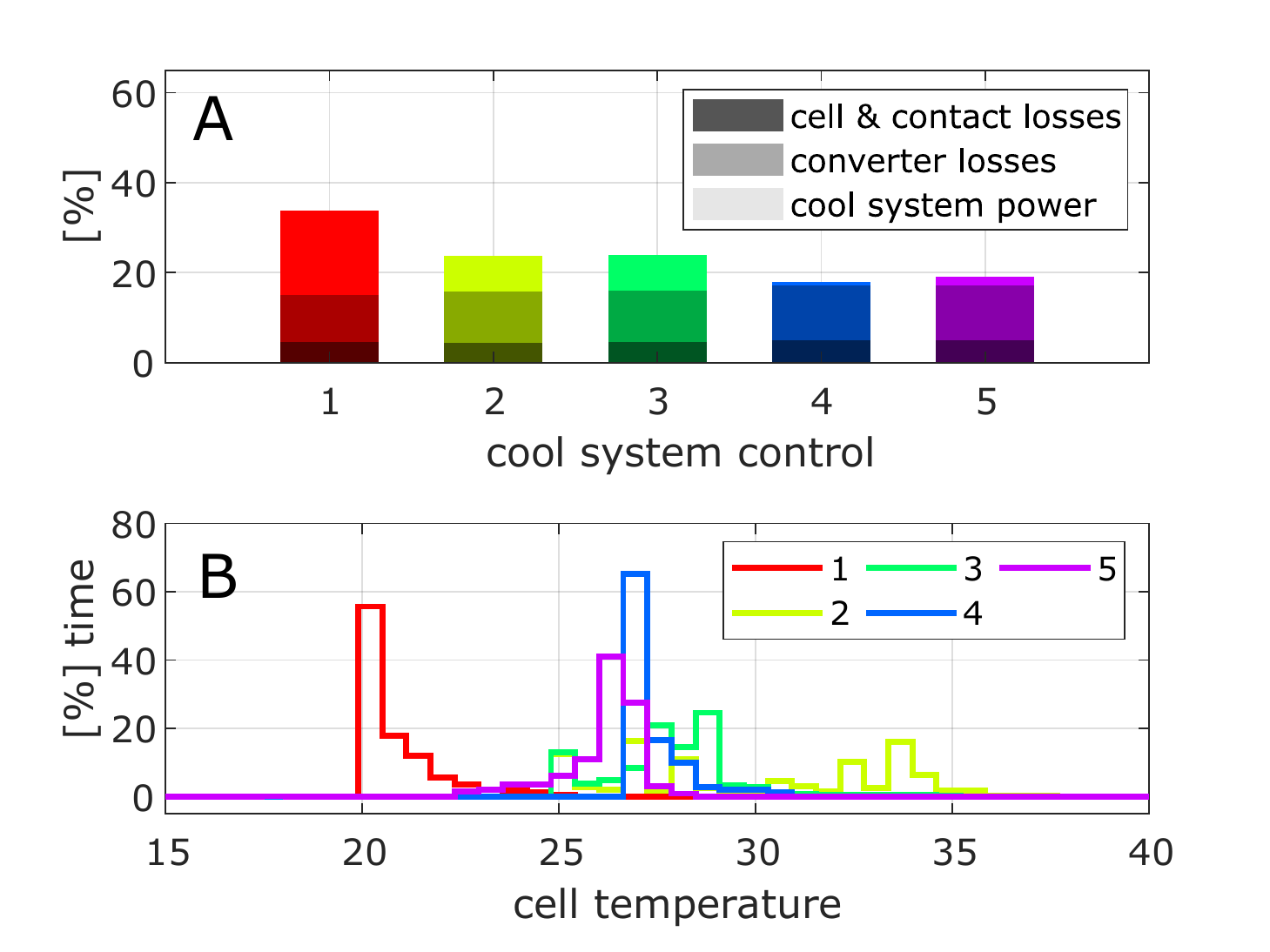}
		\caption{Thermal management system performance summary in the first week of operation. A shows the different contributions to the losses; B shows the average temperature histogram of the cells}
		\label{fig:control_etaT}
	
		% Exact numbers of each loss-source:
		% cells + converter + cooling system = total
		% 1: 4.6  + 10.4 + 18.8 = 33.8
		% 2: 4.4  + 11.4 + 7.8  = 23.7
		% 3: 4.45 + 11.4 + 7.9  = 23.9
		% 4: 4.9  + 12.2 + 0.8  = 17.9
		% 5: 5.0  + 12.2 + 1.8  = 19
		
	\end{figure}

	Fig.\ \ref{fig:control_etaT} compares the average performance of the five control approaches over the first week of operation. Fig.\ \ref{fig:control_etaT}.A shows the daily losses as fraction of the daily energy throughput (averaged over the first seven days). The losses in the cells, contact resistances, and converter are more or less identical in all cases---it can be seen that the main difference is due to differing cooling power requirements. Fig.\ \ref{fig:control_etaT}.B shows the histogram of the mean cell temperature, illustrating the thermal homogeneity that results from each control approach. % 
	As expected, losses associated with the cooling system are high if it is always working at full power (method 1), and the cells are held at low temperatures. The `on/off'-control methods (2 and 3) both have 10 \%-pts lower losses associated with the cooling system, but result in a large spread in cell temperatures, which is mainly due to the control of the AC system. When using the local temperature for control (method 2), the AC system will only operate once the entire battery has heated up, which takes time due to thermal inertia. During this time, the module fans are working at full power to cool the hot cells, but the cooling is not very effective due to the relatively warm air which is blown over the cells. In this scenario the AC system needs to work less since when it is on it removes a large amount of heat. When the control method however uses the hot-spot temperature measurements (method 3), the AC system will turn on as soon as at least one cell heats up. This ensures efficient cooling at the module level and keeps cells at lower temperatures, but reduces the effectiveness of the AC system since it is only ejecting a small amount of heat. Therefore, method 2 requires more power to operate the fans while method 3 requires more power to operate the AC system. In the end, both use a similar amount of energy but method 3 results in lower cell temperatures. %
	For similar reasons, the cells are overall cooler when using control method 5, where cooling is proportional and controlled by the hot-spot temperature, compared to method 4, where cooling is proportional and controlled by the local temperature---although the difference between these two approaches is small. Both methods result in significantly higher efficiencies than the other approaches because they often operate at partial power. Control method 5 is slightly less efficient than method 4 because the increased operating power for the AC system is not fully compensated for by the reduced operating power of the fans.
	
\begin{figure}
			\centering
			\includegraphics[width=16cm]{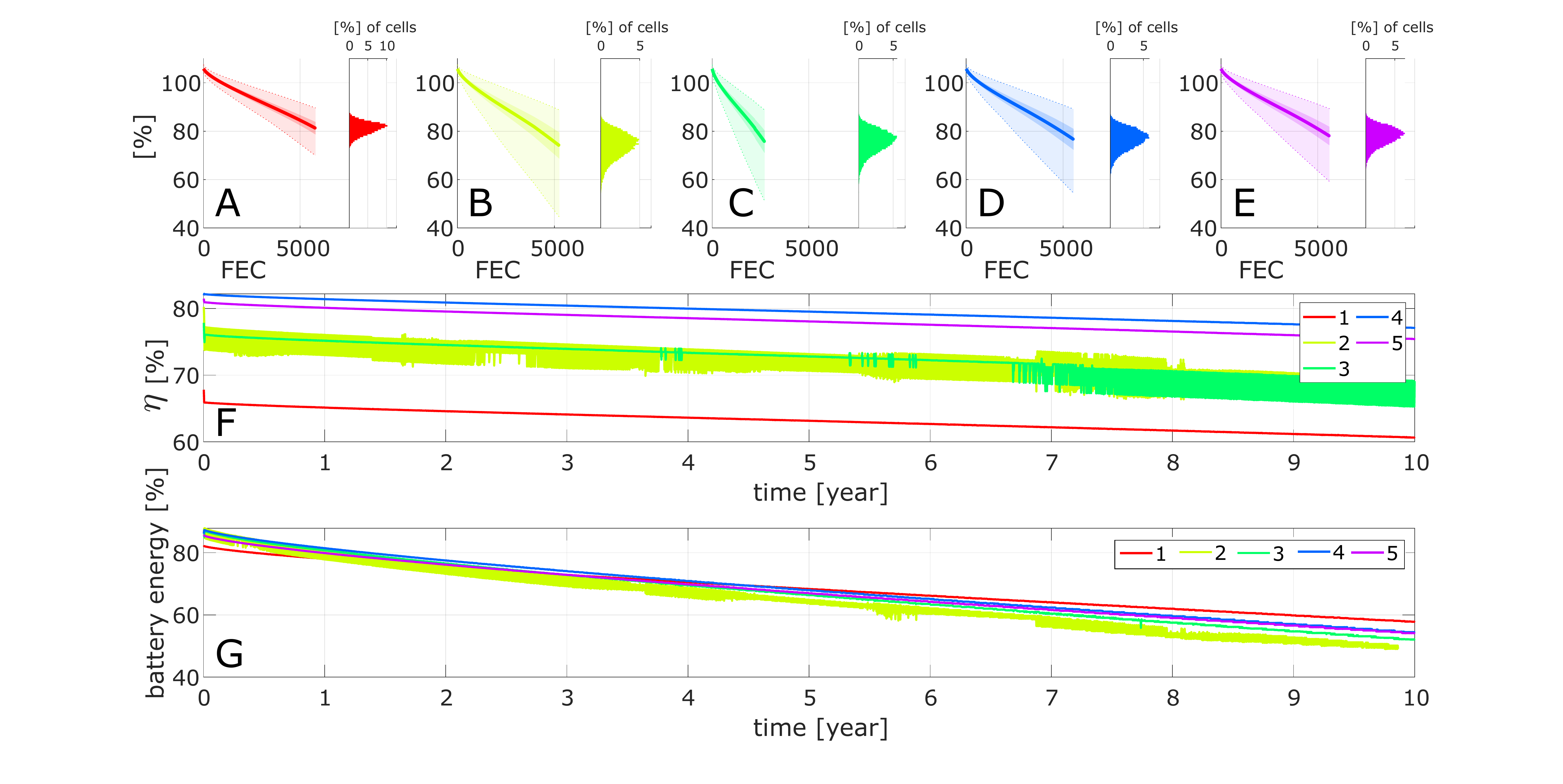}
			\caption{Impact of thermal management system approach on degradation of the battery and cells. A--E show evolution of cell charge capacities and their distributions according to each thermal control approach; F shows 1C discharge energy.}
			\label{fig:control_degrad}

		\end{figure}

	We now consider the long-term degradation impact of these various thermal management approaches, with results presented in Fig.\ \ref{fig:control_degrad}. The average temperature of the cells determines their degradation rate, which is dominated by SEI growth, while the pack temperature uniformity and current distribution determines the rate of increase of cell-to-cell variations. The top row of results, Fig.\ \ref{fig:control_degrad}A-E, shows the degradation trajectories of the cells according to the five cooling methods. Method 1, with the cooling system always at full power, results in low degradation since the cells are kept at low temperature. After 10 years, the average cell capacity with this method is 81.2\% of the nominal cell capacity, with a standard deviation of 2.6\% between cells. The `on/off'-control methods (2 and 3) result in significant temperature swings over time, causing larger cell-to-cell variations in capacity, giving respective standard deviations in capacity across the pack of 5.4\% and 4.7\% after 10 years. Method 2, `local on/off' control, results in the highest cell temperatures, often 30-\SI{35}{\celsius}, and correspondingly fast degradation to a mean capacity of 74.0\% after 10 years. Method 3, `hot-spot on/off' control, results in temperatures between the first and second methods, giving a mean capacity of 75.7\% after 10 years. %
	The proportional control methods (4 and 5), respectively based on `local' and `hot-spot' temperatures, result in cell temperatures just below those of method 3, with method 5 giving slightly lower temperatures  than method 4. After 10 years, this results in a respective mean capacities of 76.6\% and 78.0\%, with respective standard deviations of 4.4\% and 3.8\%. Note that, in the results shown in Fig.\ \ref{fig:control_degrad}, all cells experienced the same number of cycles but different numbers of full equivalent cycles due to their different degradation rates.
	%
	% mean capacities at end of 10 years:
	%   1   81.1897174933862
	%   2   73.9822480820106
	%   3   75.6710346395503
	%   4   76.5815488988095
	%   5   77.9584026851852
	%
	% STD at end of 10 years
	%   1   2.59188358820647
	%   2   5.40737049316187
	%   3   4.69806980206398
    %   4   4.34727263611763
    %   5   3.83158113985650
    %
    % Overall efficiency at start, end, difference
    % 67.8154624215420	60.6725850372817	-7.14287738426030   note initial steep drop
    % 80.2319549340633	66.2573220880796	-13.9746328459837   note alternating
    % 77.8406472706355	69.1010005196294	-8.73964675100619   note alternating
    % 81.9471141767527	77.0867772983265	-4.86033687842624
    % 81.4637760702745	75.4151776513717	-6.04859841890286
    %
    % Total usable energy at start, end, difference
    % 82.3488845988846	57.8398398398398	-24.5090447590448
    % 86.2449860574861	49.1260099385099	-37.1189761189761
    % 84.9305734305734	52.1503181753182	-32.7802552552553
    % 86.1036299692550	54.4675809738310	-31.6360489954240
    % 85.8544017231517	54.1397647647648	-31.7146369583870
    %
	
	The evolution of round-trip efficiency during long-term ageing simulations is shown in Fig.\ \ref{fig:control_degrad}.F, and as expected this decreases over time from initial values that were also given in Fig.\ \ref{fig:control_etaT}.A. All thermal control methods result in a pack that loses about 5 \%-pts round-trip efficiency after 10 years of operation. The efficiencies associated with thermal control methods 2 and 3, the on-off methods, vary from day-to-day due to thermal inertia effects causing temperature oscillations. In other words, the battery oscillates between days when the cooling system is mostly off and the system heats up,  and days when the cooling system is on for a large part of the day to cool the battery down again.
	
	The evolution of usable energy is also shown in Fig.\ \ref{fig:control_degrad}.G---here defined as the energy discharged in a full cycle after subtracting all losses that occurred in that discharge period. At the start of life, the difference in usable energy between the five thermal control methods is mostly due to the efficiency differences. The usable energy associated with thermal control method 2, `local on/off', is variable due to the alternating periods where the AC is on vs.\ off---these may be out-of-sync with the discharge cycle due to the thermal inertia of the battery. Method 5 results in a lower usable energy than method 3, but this is purely due to the time delay just described, rather than higher losses: both control methods act based on the hot-spot temperature, but method 3 uses thresholds of \SI{35}{\celsius} and \SI{30}{\celsius} respectively for the fans and the AC system, while method 5 switches on at \SI{25}{\celsius}. This means that when the discharge starts, the cooling system according to method 5 will switch on earlier, thus consuming more power during the discharge itself and reducing the usable energy. However, after the discharge, the cooling system of method 3 will still be cooling the battery unlike in method 5, but this is not factored into the usable energy. This shows how momentarily, the usable energy can be maximised by delaying action of the cooling system. However, this comes at a cost of increased degradation. %
	Since the evolution of the usable capacity over time is mostly dominated by the degradation of the cells, the distribution of the cell capacities, and the evolution of the efficiency, the usable energy resulting from thermal control method 1 is the smallest of all at the beginning of life but during the lifetime this is overtaken by the degradation caused by the other methods. 

\section{Conclusions}
This paper considered the performance of a MWh-scale grid battery using a newly developed unique large long-term simulation consisting of 18900 individual cell models coupled electrically and thermally into a system model. A comparative analysis of various aspects of system performance was undertaken by gradually adding functionality to the model. We investigated the impact of electrical contact resistances, cell-to-cell variations, and temperature variations caused by differing thermal management approaches. Despite previous studies that showed apparently large impacts on system current distribution due to high electrical contact resistances, we found increased contact resistances to have only minor impacts on the overall behaviour of a large-scale battery, where resistances are typically much lower than in small-scale lab tests or various simulation-based studies. Similarly, cell-to-cell variations in initial capacity and resistance barely affect the overall system behaviour in terms of efficiency and long-term degradation. However, variations in the rate at which individual cells degrade strongly impact the evolution of cell-to-cell variations over a the system lifetime.

We also found that lifetime and round-trip efficiency depend strongly on temperature values and uniformity, which in turn depend on the design choices made for the arrangement and control of the thermal management system. At one extreme, a system can be built to keep the cells very well cooled with minimal temperature non-uniformity, but the operating power required for this is unacceptably high---about 20\% of the total charging energy into the battery. Alternatively, on-/off-control methods require about 8\% of charging energy to operate the thermal management system but this comes at the cost of significantly increased degradation of the cells, with mean capacity decreasing by 7 \%-pts compared to the best case method, and total usable energy decreasing by 5 additional \%-pts. Control methods where the thermal management system can operate at partial power, offering cooling proportional to the temperature difference above nominal, are a good compromise and give good round-trip efficiency and lifetime. With these approaches only a small amount of the overall energy is needed to operate the thermal management system, and the additional system degradation is limited to 3 \%-pts of the overall battery compared to the first case.

Overall this work shows the critical impact on grid battery long-term performance of considering the interaction between cell behaviour and system design, especially with respect to the design and control of the thermal management system.

\section*{Acknowledgements}
This work is part of the Energy Superhub Oxford project funded by InnovateUK (grant ref.\ 104779). %For the purpose of Open Access, the authors apply a CC BY public copyright licence to any Author Accepted Manuscript (AAM) version arising from this submission.
%\section*{References}

\bibliographystyle{elsarticle-num}
\bibliography{References}

\end{document}